\documentclass[preprint,aps,showpacs,preprintnumbers,amsmath,amssymb,nofootinbib]{revtex4}
\usepackage{epsfig}
\usepackage{graphicx}

\begin{document}

\begin{flushright}
\end{flushright}


\newcommand{\be}{\begin{equation}}
\newcommand{\ee}{\end{equation}}
\newcommand{\bea}{\begin{eqnarray}}
\newcommand{\eea}{\end{eqnarray}}
\newcommand{\nn}{\nonumber}
\def\CP{{\it CP}~}
\def\cp{{\it CP}}

\title{\large Towards extracting the best possible results from NO$\nu$A}
\author{Soumya C. and R. Mohanta }
\affiliation{School of Physics, University of Hyderabad, Hyderabad - 500 046, India }

\begin{abstract}
The  NuMI Off-Axis $\nu_{e}$ Appearance (NO$\nu$A) is the currently running leading long-baseline neutrino oscillation experiment, 
whose main physics goal is to explore the current issues in the neutrino sector, such as determination of the  neutrino mass ordering, 
resolution of the  octant of atmospheric mixing angle and to constrain the Dirac-type CP violating phase $\delta_{CP}$. 
In this paper, we  would like to investigate whether it is possible to extract the best possible results from NO$\nu$A with a 
shorter time-span than its scheduled run period by analyzing its capability to discriminate the degeneracy among  various 
neutrino oscillation parameters 
within four years of run time, with two years in each neutrino and antineutrino modes. Further, we study the same by adding the data 
from T2K experiment for a total of five years run with 3.5 years in neutrino mode and 1.5 years in antineutrino mode. 
We find that NO$\nu$A (2+2) has a better  oscillation parameter degeneracy discrimination capability  compared to its scheduled run
period for four years,  i.e, NO$\nu$A (3+1).
\end{abstract}

\maketitle

\section{Introduction}

The results from various neutrino oscillation experiments \cite{exp-1, exp-2,exp-3,exp-4,exp-5,exp-6,exp-7,exp-8} confirm that neutrino flavours mix with each other and neutrinos 
do possess tiny but non-zero masses. This mixing of neutrino can be described by a unitary matrix, so called Pontecorvo-Maki-Nagakawa-Sakata (PMNS) matrix, 
which is parameterized by three mixing angles, often referred to as the solar mixing angle ($\theta_{12}$), atmospheric mixing angle ($\theta_{23}$), reactor mixing angle ($\theta_{13}$)
 and a Dirac-type CP violating phase ($\delta_{CP}$) \cite{pmnsa1,pmnsa2}. The probability of neutrino oscillation  depends on these parameters as well as 
on two mass squared differences namely, the solar mass squared difference ($\Delta m^2_{21}$) and the atmospheric mass squared difference ($\Delta m^2_{31}$). 
All these parameters are determined through various neutrino experiments except the Dirac CP phase. However, we do not know the mass ordering of neutrinos, 
i.e, the sign of $\Delta m^2_{31}$ and this left us with two choices like normal ordering/hierarchy (NH) with $\Delta m^2_{31}>0$ and inverted ordering/hierarchy 
with $\Delta m^2_{31}<0$. Furthermore, recent experimental result from MINOS \cite{minos23} shows that $\theta_{23}$ is non-maximal. Therefore,  
 the octant of the mixing angle $\theta_{23}$ remains unknown, i.e, whether  it lies in lower octant (LO) i.e., $\theta_{23}<45^\circ$ or  
in higher octant (HO) with  $\theta_{23}> 45^\circ$. There are many neutrino oscillation experiments which are intended to determine these unknowns.
Among all these experiments, NO$\nu$A is one of the new generation accelerator based long-baseline experiment which aims to determine most of these 
unknown parameters.

NO$\nu$A \cite{nova} experiment is  currently running long-baseline neutrino oscillation experiment, which uses an upgraded NuMI beam power 
of 0.7 MW at Fermilab. It has a 14 kton totally active scintillator  detector (TASD) placed $0.8^\circ$ off-axis from the NuMI beam near the Ash River, 
situated 810 km far away from Fermilab. It also has a 0.3 kton near detector located at
the Fermilab site to monitor the un-oscillated neutrino or anti-neutrino flux.
This experiment is designed to observe both $\nu_e (\bar{\nu}_e)$ appearance events and  $\nu_{\mu} (\bar{\nu}_{\mu})$ disappearance events. 
The main physics goals of this experiment are
\begin{itemize}
  \item Appearance events: To determine the value of $\theta_{13}$, determination of the octant of $\theta_{23}$, mass ordering and constrain the Dirac CP phase.
  \item Disappearance events: The precision measurement of atmospheric oscillation parameters, $\Delta m^2_{23}$ and $\theta_{23}$.
\end{itemize}

The determination of these parameters by an oscillation experiment like NO$\nu$A, which is mainly rely on the oscillation probability, is extremely 
difficult due to the parameter degeneracies, since various combination of these parameters give the same probability. A lot of work has been done 
in the literature to resolve these degeneracies among oscillation parameters \cite{degen-1,degen-2,degen-3}. Moreover, there was a suggestion
for the need of an early anti-neutrino run to get a first hint of mass ordering in NO$\nu$A \cite{suprabh}.
There it  has been shown that the sensitivity for the determination of mass hierarchy is above 2$\sigma$ (i.e., $\chi^2>4$) only for $\delta_{CP}$ 
value around $\mp 90^\circ$ for true hierarchy and octant as NH-LO or HO-IH, where the scheduled run time, i.e., (3 yrs in $\nu$ mode + 0 yr in $\bar \nu$ mode)
 gives almost
null sensitivity. 
The scheduled  run period of NO$\nu$A is  for a total of six years with first three years in neutrino mode followed by the next three years in antineutrino mode. 
Therefore, it is of great importance to study the ability to discriminate the degeneracies between different oscillation parameters of this experiment 
within a minimal time-span, since it leads to an early  understanding of neutrino oscillation parameter space. In this context,  
we investigate in this paper how to extract the  best possible  results from NO$\nu$A with shortest time-span by analyzing  its   physics potential 
and degeneracy discrimination capability for a total of four years of runs, with two years in each neutrino and antineutrino modes. 
We have shown that the (2 + 2) years of run will provide  much better sensitivity for the mass hierarchy
determination in comparison to its  scheduled run  for four years i.e, 3 years in neutrino mode followed by one year in anti neutrino mode.  
Furthermore, we study the same by adding data from T2K experiment for a total of five years run with 3.5 years in neutrino mode and 1.5 years 
in antineutrino mode.\\

The outline of this paper is as follows. In section II, we present the  details of simulation of T2K and NO$\nu$A experiments that we have  considered in this work. 
The neutrino  oscillation parameter degeneracies are discussed in section section III. Section IV contains the  discussion about the mass ordering and octant 
determination. Finally, we summarize our results in section V.

\section{Simulation details}
We simulate the neutrino oscillation events for T2K (Tokai-to Kamioka) as well as NO$\nu$A experiments by using  GLoBES package \cite{Huber:2004-1,Huber:2009-2}. 
T2K is also a currently running off-axis long-baseline experiment, which has been designed to study the phenomenon of neutrino oscillation. 
It uses an upgraded beam power of 0.77 MW 
and  has the water cherenkov detector of mass 22.5 kton placed about 295 km away from Tokai.
We simulate T2K experiment with updated experimental description as given in \cite{t2k-5}. As we mentioned earlier, NO$\nu$A experiment is an off-axis 
experiment with a baseline of 810 km, which uses a beam power of 0.7 MW and a detector of mass 14 kton. The experimental specifications of NO$\nu$A are taken 
from \cite{ska12} with the following characteristics:\\
 Signal efficiency: 45\% for $\nu_{e}$ and $\bar{\nu}_{e}$ signal; 100\% $\nu_{\mu}$ CC and $\bar{\nu}_{\mu}$ CC.\\
Background efficiency: \\
a) Mis-ID muons acceptance: 0.83\% for $\nu_{\mu}$ CC, 0.22\% for $\bar{\nu}_{\mu}$ CC ; \\
b) NC background acceptance: 2\% for  $\nu_{\mu}$ NC,  3\% for $\bar{\nu}_{\mu}$ NC ; \\
c) Intrinsic beam contamination: 26\% for $ \nu_e$,  18\% for $\bar{\nu}_{e}$\;, \\
and we consider $5 \%$ uncertainty on signal normalization and $10 \%$ on background normalization.
The migration matrices for NC background smearing are taken from \cite{ska12}.
 The true values of oscillation parameters that we use in our simulation are listed in the Table-\ref{truevalues} \cite{expt-data}.

\begin{table}[h]
\begin{center}
\vspace*{0.1 true in}
\begin{tabular}{|c|c|}
\hline
 $\sin^2\theta_{12}$ & 0.32 ~ \\
\hline
$\Delta m_{21}^2$ & $7.6 \times 10^{-5}~ {\rm eV}^2$ ~ \\
\hline
 $\sin^2 2\theta_{13}$ & 0.1 ~ \\
\hline
$\Delta m_{atm}^2$ & $2.4 \times 10^{-3} ~{\rm eV}^2$ for NH ~ \\
                                & $-2.4 \times 10^{-3} ~{\rm eV}^2$ for IH ~ \\
\hline
 $\sin^2 \theta_{23}$ & 0.41 (LO), 0.59 (HO) ~ \\
\hline
$\delta_{CP} $ & $0^\circ$  ~ \\
\hline
\end{tabular}
\end{center}
\caption{The true values of oscillation parameters considered in the simulations taken from \cite{expt-data}.}
\label{truevalues}
\end{table}


\begin{figure}[!htb]
\label{biprobability}
\includegraphics[width=7cm,height=7cm]{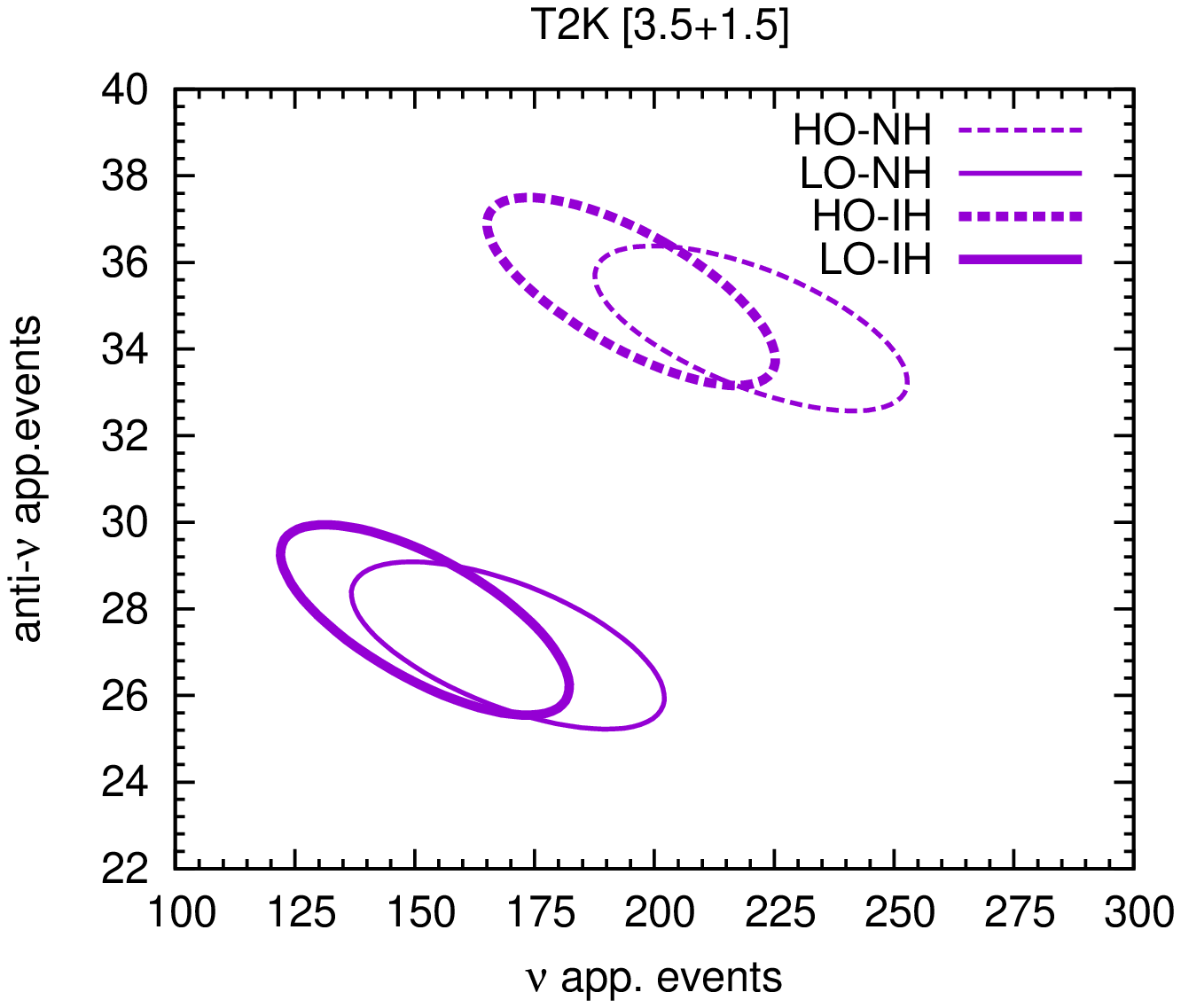}
\includegraphics[width=7cm,height=7cm]{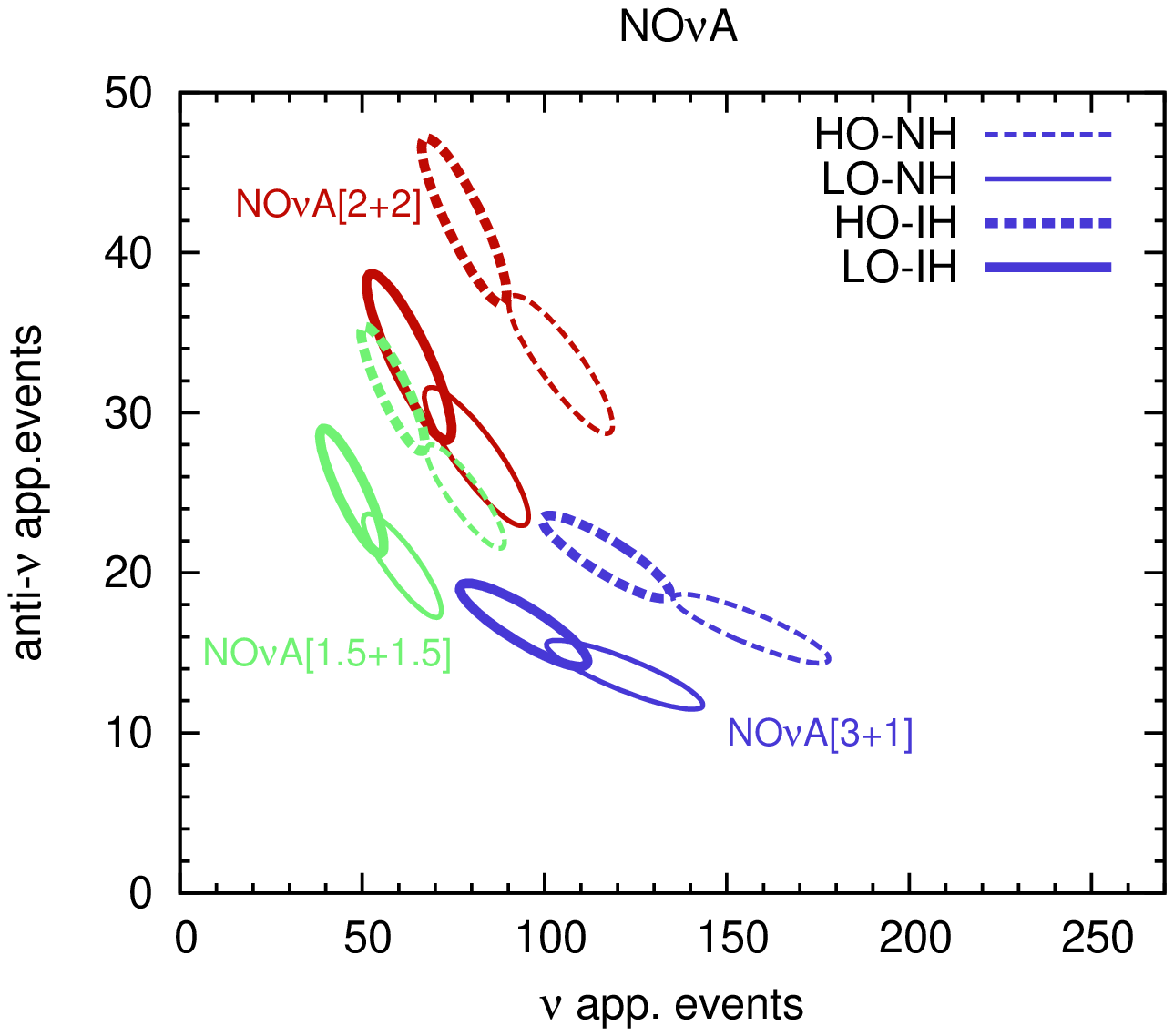}\\
\caption{Neutrino and antineutrino appearance events for the $\nu_{\mu} \rightarrow \nu_{e}$ versus $\bar \nu_{\mu} \rightarrow \bar\nu_{e}$ channels  by assuming both IH and NH and for lower and higher octants of $\theta_{23}$.}
\end{figure}


\section{Neutrino oscillation parameter degeneracies}
The parameter degeneracies in neutrino oscillation sector are of mainly three types and they are: ($\delta_{CP},\theta_{13}$), 
sign of $\Delta m^2_{31}$ and ($\theta_{23},\pi/2-\theta_{23}$). Recently, the reactor experiments such as Daya Bay \cite{daya-bay,daya-bay1}, 
Double Chooz \cite{doublecz} and RENO \cite{reno} have precisely measured the value of the reactor angle as  $\sin^22\theta_{13}\approx 0.089\pm 0.01$. 
Therefore, the eight fold degeneracy is reduced to four-fold degeneracy.  
Out of these degeneracies, the degeneracy in which $\theta_{23}$ can't be distinguished from $(\pi/2-\theta_{23})$ is called octant degeneracy and 
the degeneracy in the sign of $\Delta m^2_{31}$  is called hierarchy ambiguity. So far, we are left with four degeneracies and they are 
represented  as NH-HO, NH-LO, IH-HO and IH-LO, where NH/IH (HO/LO) stands for Normal/Inverted  ordering (Higher/Lower Octant). 
Resolution of these degeneracies are the main challenges of the present and future  long-baseline neutrino oscillation experiments, 
which are mainly looking for 
oscillation from $\nu_{\mu}(\bar\nu_{\mu}) \rightarrow \nu_{e} (\bar\nu_{e})$. The expression for the oscillation probability, which is up to 
first order in $\sin \theta_{13}$ and $\alpha \equiv \Delta_{21}/\Delta_{31}$ is given as \cite{akhmedov,cervera,freund}
\begin{eqnarray}\label{prob}
P(\nu_\mu\to\nu_e) & \approx & \sin^22\theta_{13}\sin^2\theta_{23}\frac{\sin^2(\hat A-1)\Delta}{(\hat A-1)^2} \nn\\
                & +&\alpha\cos\theta_{13} \sin 2\theta_{12}\sin2\theta_{13} \sin 2 \theta_{23}\frac{\sin \hat A\Delta}{\hat A}
\frac{\sin(\hat A-1)\Delta}{(\hat A-1)}\cos(\Delta+\delta_{CP})\;,
\end{eqnarray}
 where $\Delta =\Delta_{31}L/4E$ with $\Delta_{ij}=m_i^2-m_j^2$ and   $\hat A = 2 \sqrt 2 G_F n_e E/\Delta_{31}$, where $G_F$ is  the Fermi coupling constant 
and $n_e$ is the electron number density. For neutrinos, $\hat A$ is positive for NH and negative for IH. For antineutrino, $\hat A$ and 
$\delta_{CP}$ reverse their sign, i.e,  $\hat A \rightarrow  -\hat A$  and  $\delta_{CP} \rightarrow -\delta_{CP}$ for $P(\bar\nu_\mu\to\bar\nu_e)$.

The best way to express the degeneracies without any mathematical expression is simply by using bi-events curves. The bi-events plots for various octant-hierarchy 
combinations of  T2K and NO$\nu$A are depicted in  Fig. 1, which are obtained by computing the $\nu(\bar\nu)$ appearance events 
for the  full range of $\delta_{CP}$ with a particular octant-hierarchy combination. From the plots, we can see that the ellipses for two hierarchies 
overlap with each other for both T2K and NO$\nu$A for Lower Octant, which show that they have poor mass hierarchy discrimination capability. Whereas, 
the overlap is minimal for Higher Octant in the case of NO$\nu$A, which shows that NO$\nu$A has better degeneracy discrimination capability  compared to T2K. 
However, the ellipses for 
HO and LO are very well separated and they have good octant resolution capability. Moreover, NO$\nu$A (2+2) has better capability to determine the 
octant of $\theta_{23}$ among all other combinations due to balanced  $\nu$ and $\bar\nu$ runs.   

\section{Mass hierarchy and Octant determination}
In this section, we obtain the potential of NO$\nu$A experiment to determine the mass hierarchy and  octant of  atmospheric mixing angle and discuss 
the role of mass hierarchy-octant parameter degeneracy in the determination of these parameters.
 
\subsection{Mass hierarchy determination}
For the mass hierarchy determination, we obtain the sensitivity by calculating the  $\chi^2$  with which one can rule out the wrong hierarchy 
from the true hierarchy. We express this sensitivity as a function of true value of $\delta_{CP}$, since  it can be seen from Eq. (\ref{prob}) that 
there exists a  degeneracy between hierarchy and $\delta_{CP}$. Therefore, we simulate true events by taking NH (IH) as true hierarchy and test 
events by taking IH (NH) as test hierarchy for each true value of $\delta_{CP}$. We obtain the  $\chi^2$ by using  GLoBES and compare both event 
rates for full range of $\delta_{CP}$.  We do marginalization over all other parameters in order to get minimum $\chi^2$. We also add a prior on 
$\sin^22 \theta_{13}$. We obtain this $\chi^2$ for various true values of $\sin^2 \theta_{23}$ (i.e, $\sin^2 \theta_{23}=0.5$ for maximal mixing and 
$\sin^2 \theta_{23}=0.41~(0.59)$ for LO (HO), since the octant of atmospheric mixing angle is unknown.

\begin{figure}[!htb]
\includegraphics[width=7cm,height=7cm]{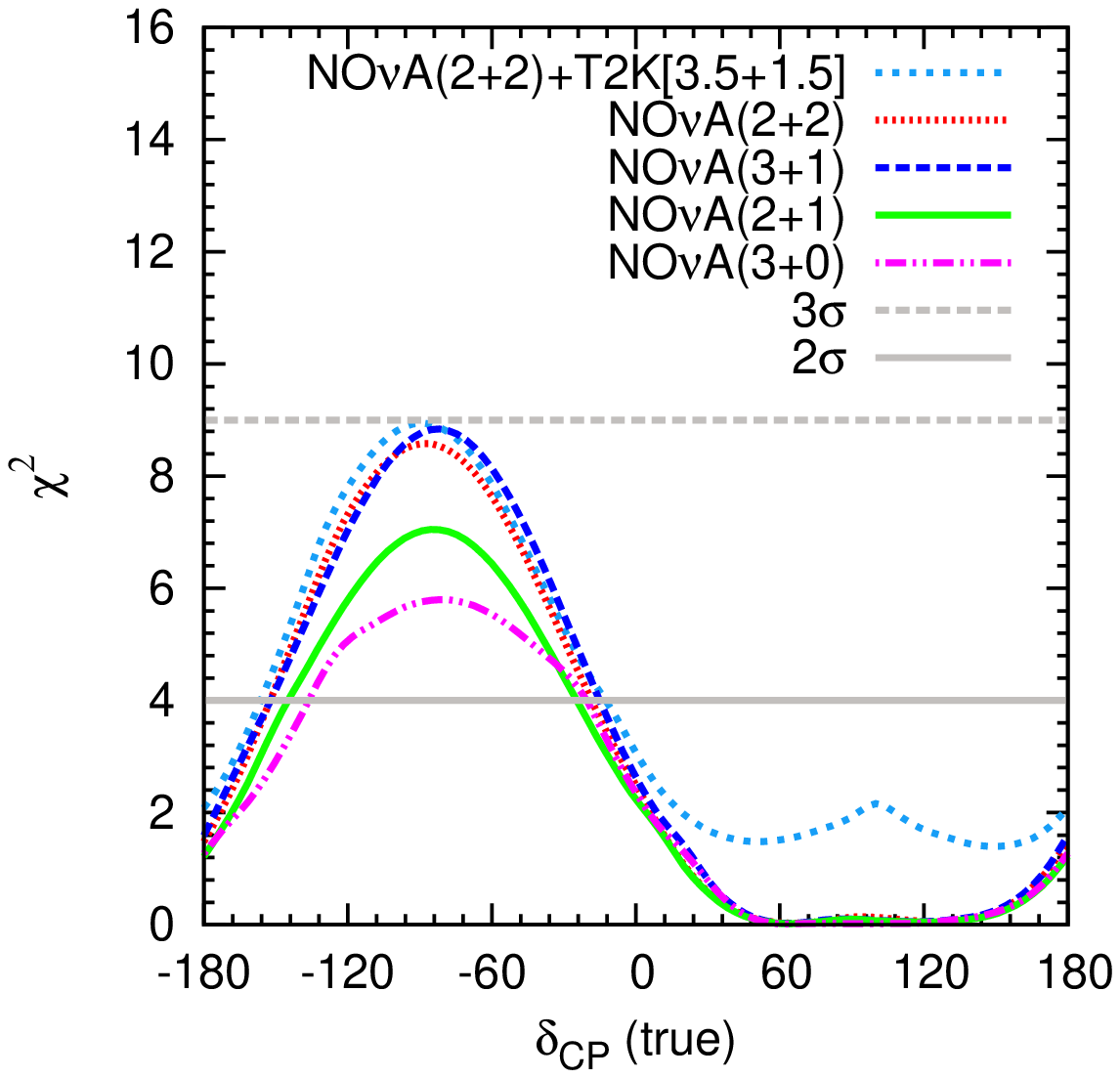}
\includegraphics[width=7cm,height=7cm]{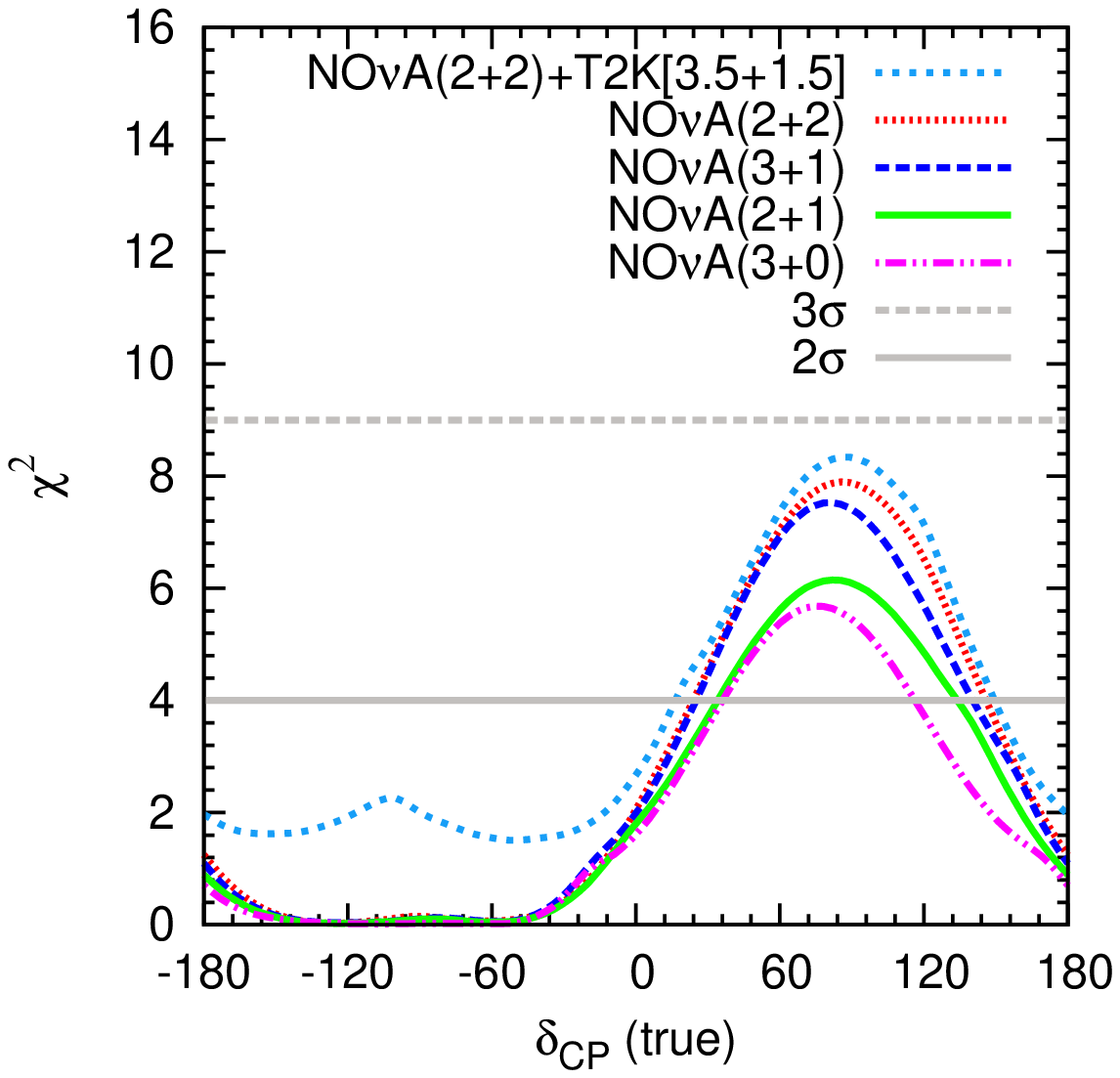}\\
\caption{The potential of determination of mass hierarchy. NH is considered as true hierarchy in the left panel and IH considered as 
true hierarchy in right panel.}
\label{MH-maximal}
\end{figure}

\begin{figure}[htb]
\includegraphics[width=7cm,height=7cm, clip]{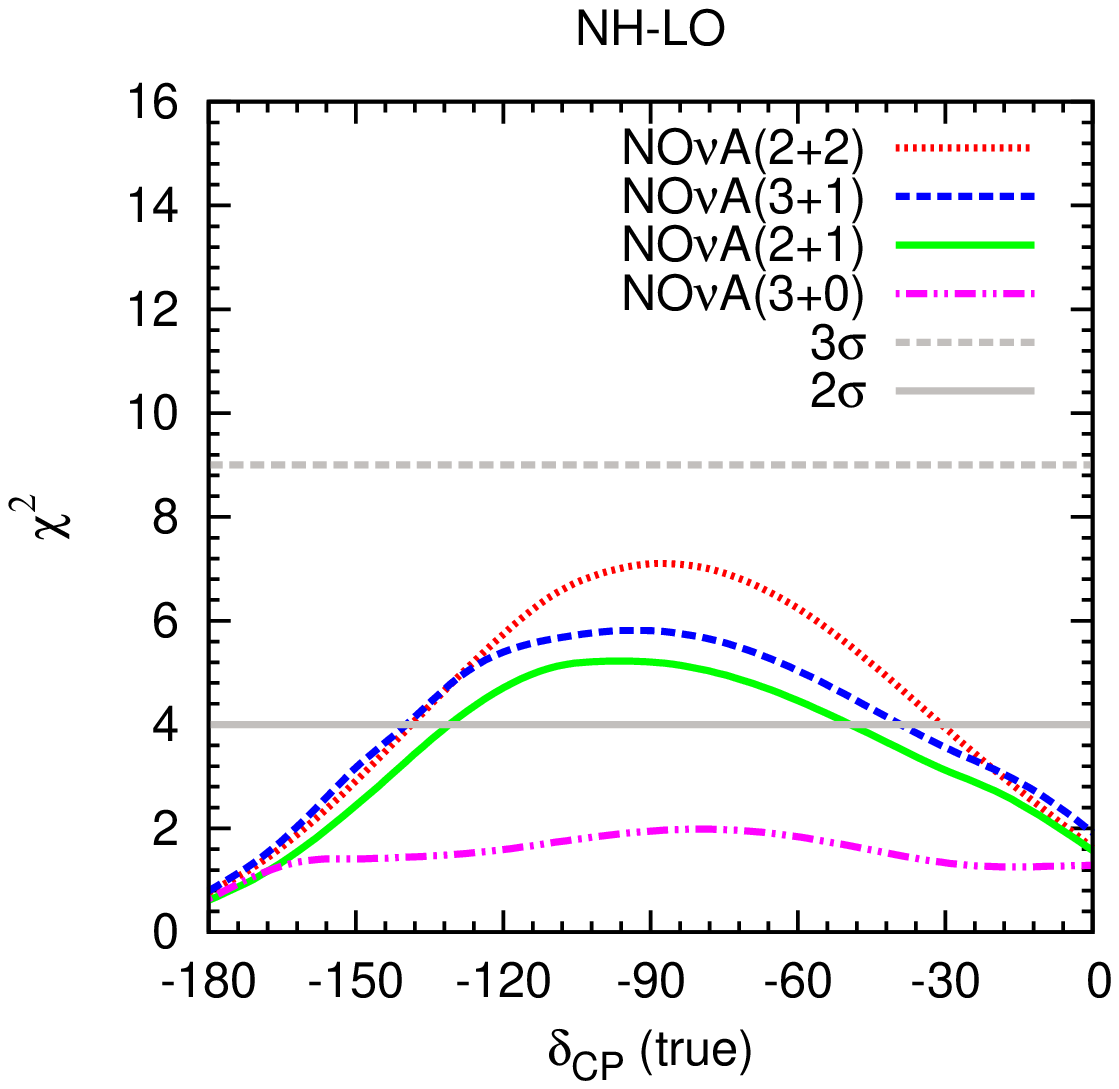}
\hspace{0.2 cm}
\includegraphics[width=7cm,height=7cm, clip]{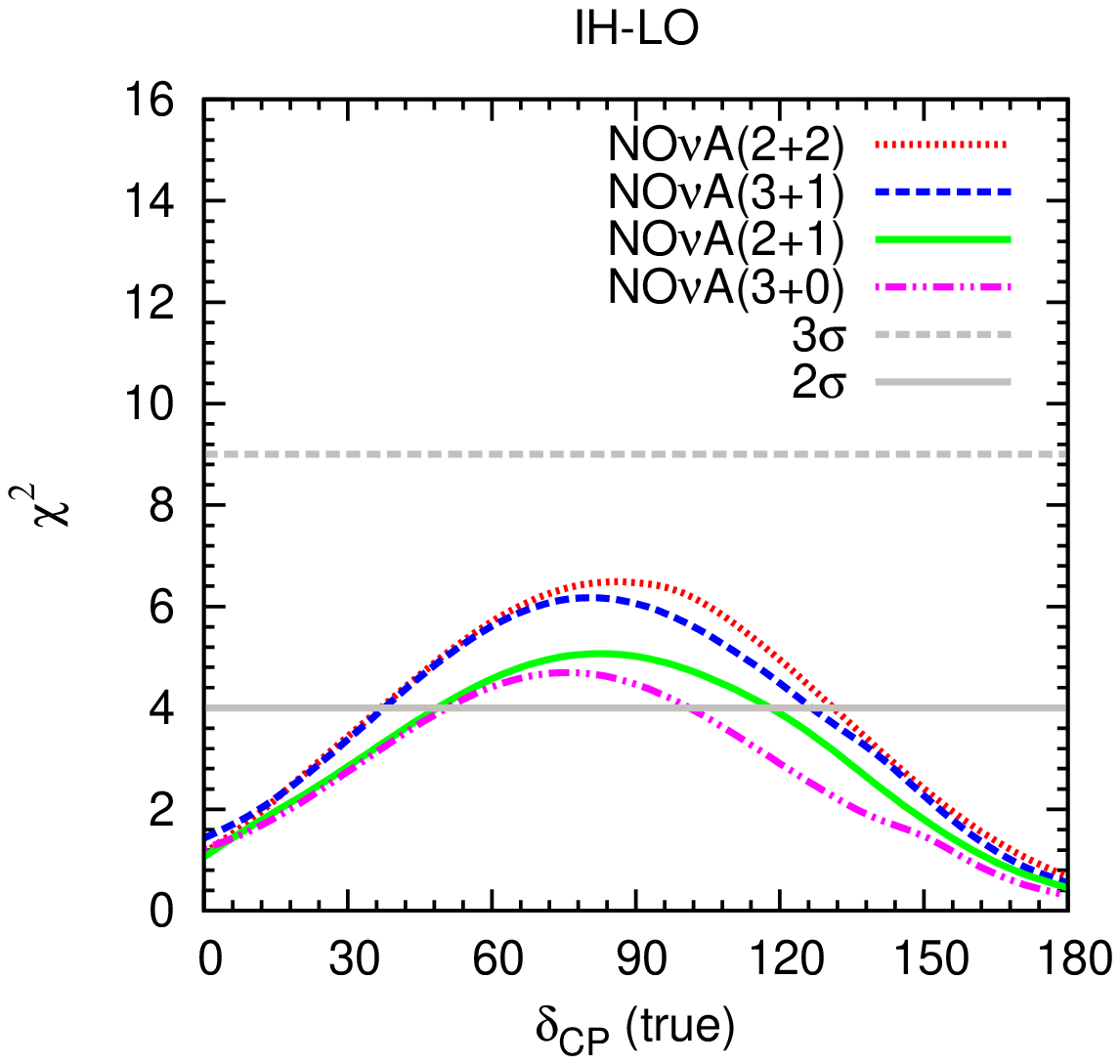}
\includegraphics[width=7cm,height=7cm, clip]{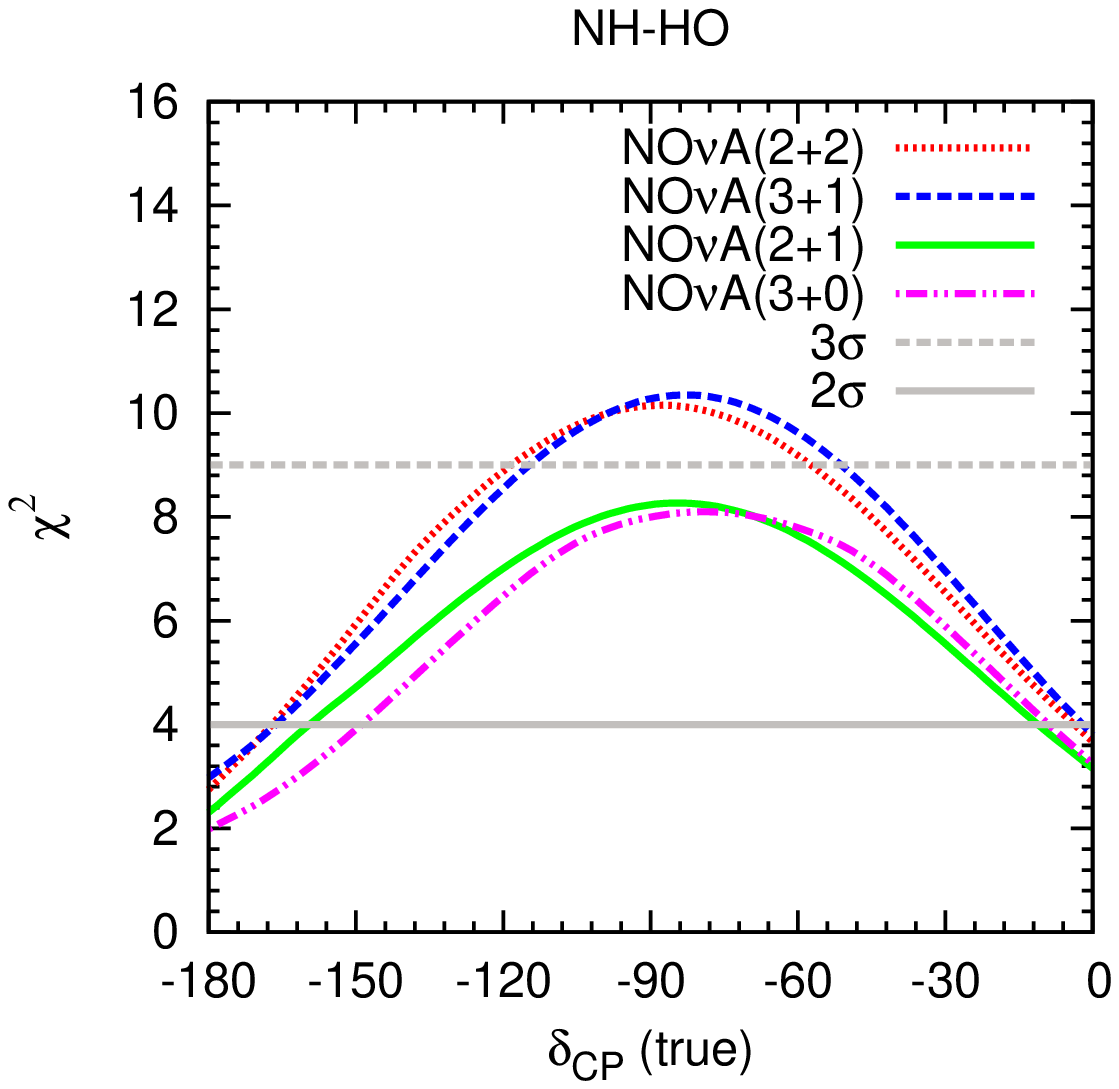}
\hspace{0.2 cm}
\includegraphics[width=7cm,height=7cm, clip]{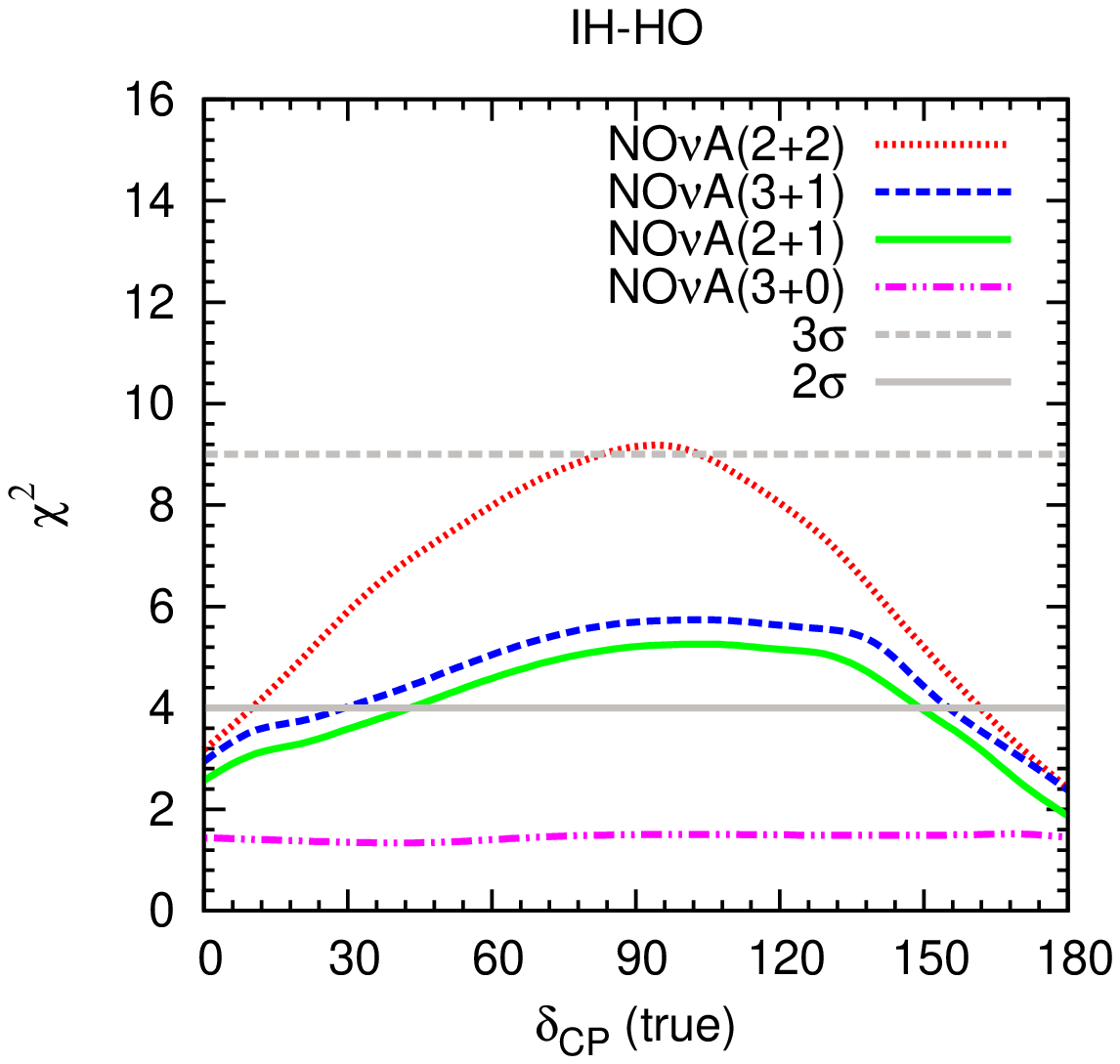}
\caption{The mass hierarchy sensitivities. The top (bottom) panel is for LO (HO), where we have used
 $\sin^2 \theta_{23}=0.41 ~(0.59)$ for LO (HO) and left (right) panel is for true NH (IH).}
 \label{MH-non-maximal}
\end{figure}

In Fig. \ref{MH-maximal}, we plot the value of $\chi^2$, obtained  for maximal mixing of atmospheric angle, as a function of $\delta_{CP}$. 
The left panel corresponds to true NH and the right panel is for true IH. From these figures, we can see that the potential to determine mass 
hierarchy for NO$\nu$A  is above 2$\sigma$ for less than half of parameter
space of $\delta_{CP}$  and it also depends on the neutrino mass ordering. The mass hierarchy sensitivity  of NO$\nu$A (2+2) is lower (higher) than that of NO$\nu$A (3+1) for 
true NH (IH) and maximal mixing of atmospheric mixing angle. The sensitivity  increases for a combined analysis of NO$\nu$A (2+2) 
and T2K (3.5+1.5) and has a 3$\sigma$ significance in the case of true NH. The mass hierarchy sensitivities for non-maximal atmospheric 
mixing angle are presented in Fig. \ref{MH-non-maximal}. We consider all possible combinations with $\sin^2 \theta_{23}=0.41 ~(0.59)$ for 
LO (HO) for  different combinations of neutrino and antineutrino run modes like NO$\nu$A (2+1), NO$\nu$A (2+2), NO$\nu$A (3+0) 
and NO$\nu$A (3+1). From these plots, we can see that the value of $\chi^2$ is always above 6 for all cases of NO$\nu$A (2+2),
 whereas for NO$\nu$A (3+1) the $\chi^2$ value is below 6 ($\sim 2.4 \sigma $) for two combinations (NH-LO and IH-HO). Hence, NO$\nu$A (2+2) has a good mass 
hierarchy discrimination capability compared to the scheduled run of NO$\nu$A for four years. Thus, we can have an early information about
the nature of mass ordering if NO$\nu$A runs in $(2\nu+2 \bar{\nu})$ mode rather than its scheduled run of $(3\nu+1 \bar{\nu})$ years. Furthermore,
if nature would be kind enough in the sense that the real mass ordering is inverted in nature and $\theta_{23}$ lies in the higher octant, 
then mass hierarchy  can be determined with more than $2\sigma$ C.L.
for  values of $\delta_{CP}$ in the range  $[0:180]^\circ$ with $(2\nu+2 \bar{\nu})$ years of run. Also if we compare the results of  three years of run,  
the sensitivity for
 the determination of mass hierarchy is better for (2+1) combination than the scheduled (3+0) combination. This in turn implies that 
there would be better perspective if NO$\nu$A runs in antineutrino mode after completing two years of run in neutrino mode.

\subsection{Octant of $\theta_{23}$ determination}
The hint of non-maximal atmospheric mixing angle observed by the MINOS Collaboration \cite{minos23} is one of the recent subject of interest in neutrino oscillation sector. 
The deviation of $\theta_{23}$ from maximal ends up with two solutions so called  lower octant ($\sin^2\theta_{23}<0.5$) and 
higher octant ($\sin^2\theta_{23}>0.5 $). For the determination of resolution of octant of $\theta_{23}$, we obtain the minimum $\chi^2$ 
with which one can rule out the wrong octant from the true octant. Therefore, we simulate true events by taking LO (HO) as true octant and 
test events by taking HO (LO) as test octant. In order to obtain the  $\chi^2$, we compare  the true events and test events for  true values of 
$\sin^2\theta_{23}$ in the range [0.32:0.68]. We marginalize over other parameters $\sin^22\theta_{13}$, $\Delta m^2_{31}$ within their 
3$\sigma$ range and $\delta_{CP}$ in its full range.  We also add a prior on $\sin^22\theta_{13}$.

\begin{figure}[!htb]
\includegraphics[width=7cm,height=7cm]{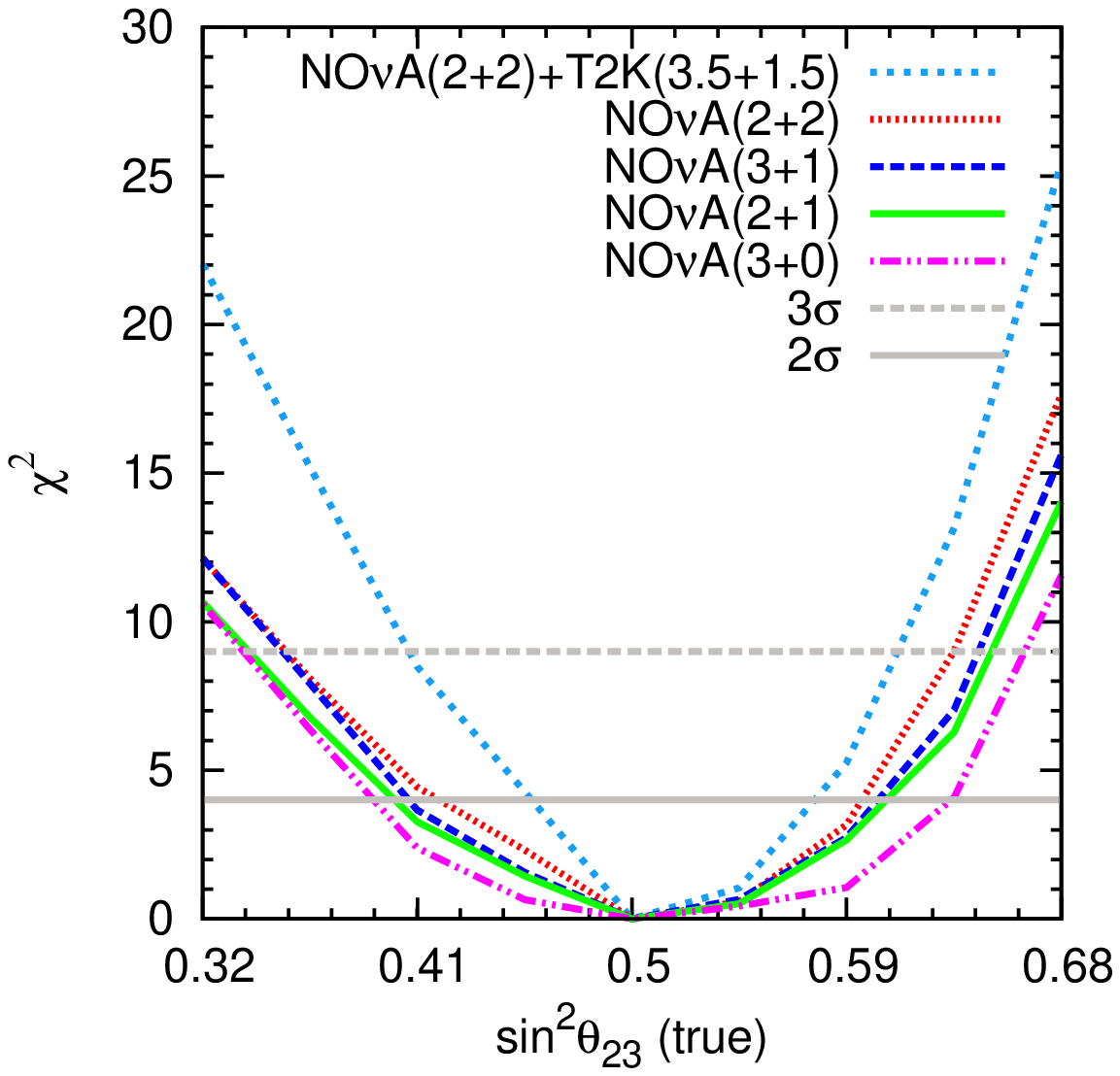}
\includegraphics[width=7cm,height=7cm]{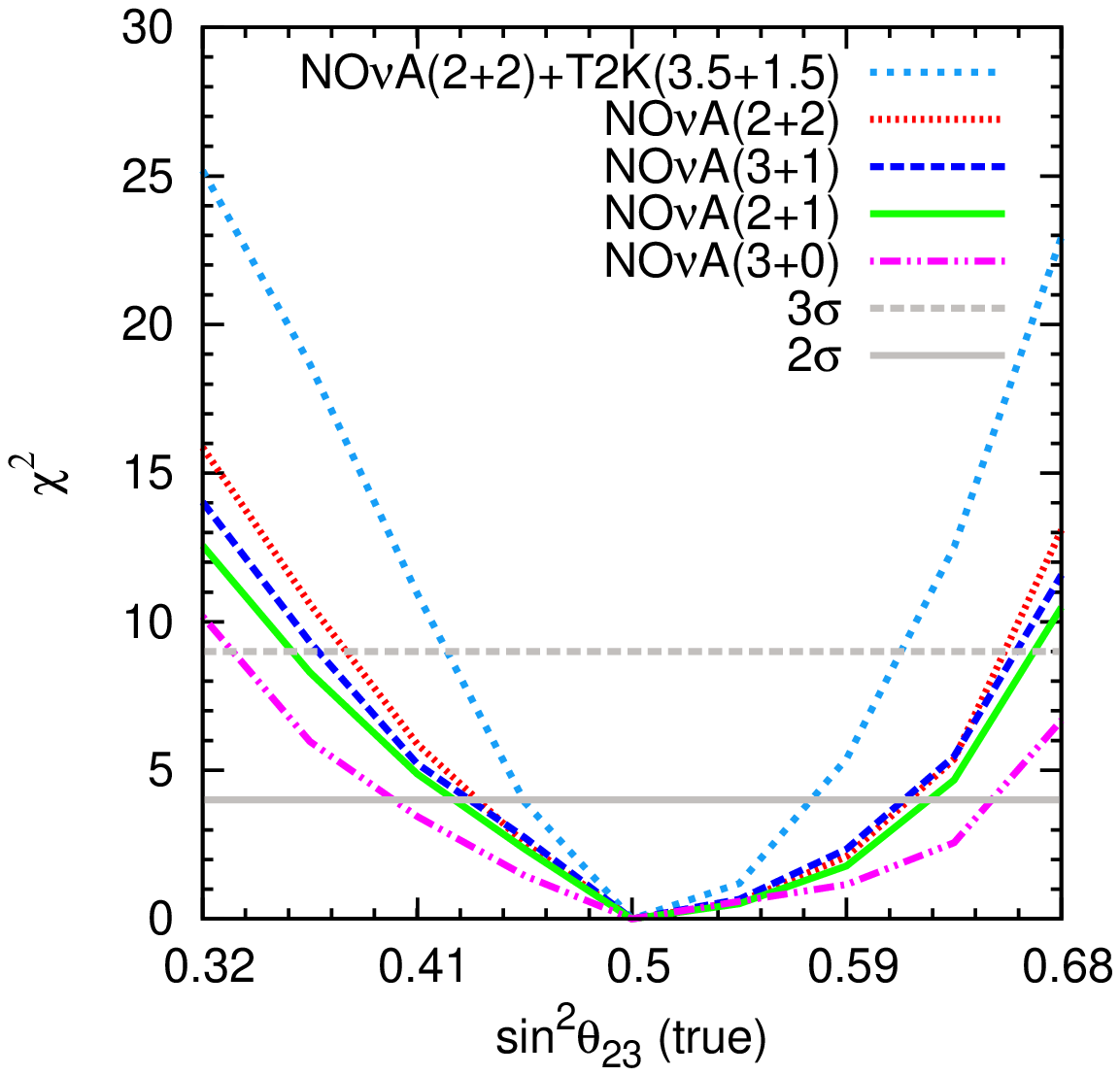}\\
\caption{The potential of octant resolution. NH is considered as true hierarchy in the left panel and IH considered as true hierarchy in right panel.}
\label{octant}
\label{MH-nonmaximal}
\end{figure}
 In Fig. \ref{octant}, the obtained  $\chi^2$ is plotted as a function of $\sin^2\theta_{23}$. Left panel corresponds to NH and the right panel 
corresponds to IH as true hierarchies. From the plots, it is clear that the potential to determine the octant of atmospheric angle is better for NO$\nu$A (2+2), 
when compared with NO$\nu$A (3+1). We can also see that a combined analysis of NO$\nu$A (2+2) and T2K (3.5+1.5) has good  octant resolution sensitivity.

\subsection{Correlation between $\theta_{23}$ and $\Delta m^2_{32}$}

The  discovery reach of mass hierarchy and octant of atmospheric mixing angles are crucial  because of the degeneracies between the oscillation 
parameters. Therefore, resolution of these degeneracies is very important  to have a clear understanding of the neutrino mixing phenomenon. 

\begin{figure}[!htb]
\includegraphics[width=7cm,height=7cm]{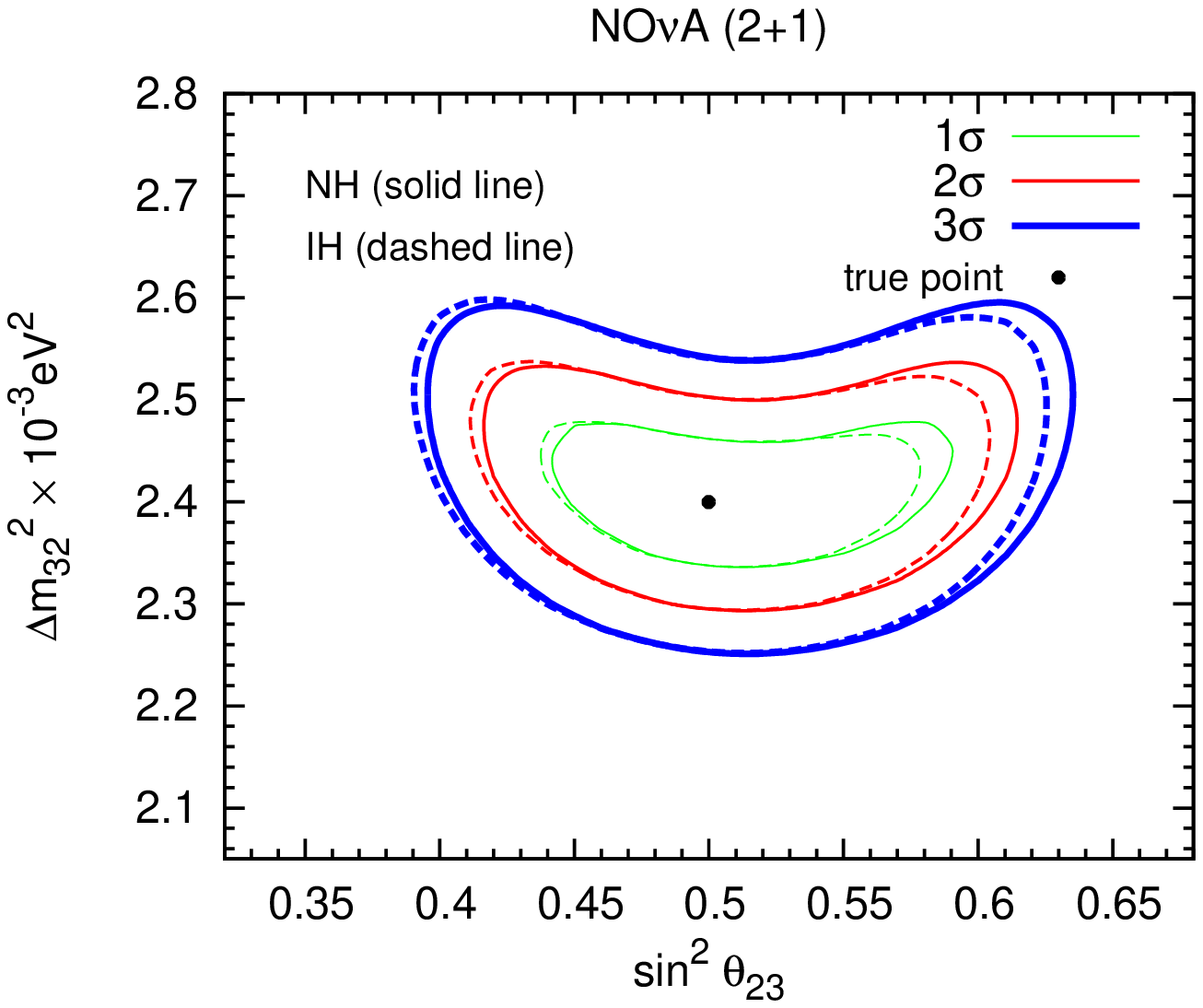}
\includegraphics[width=7cm,height=7cm]{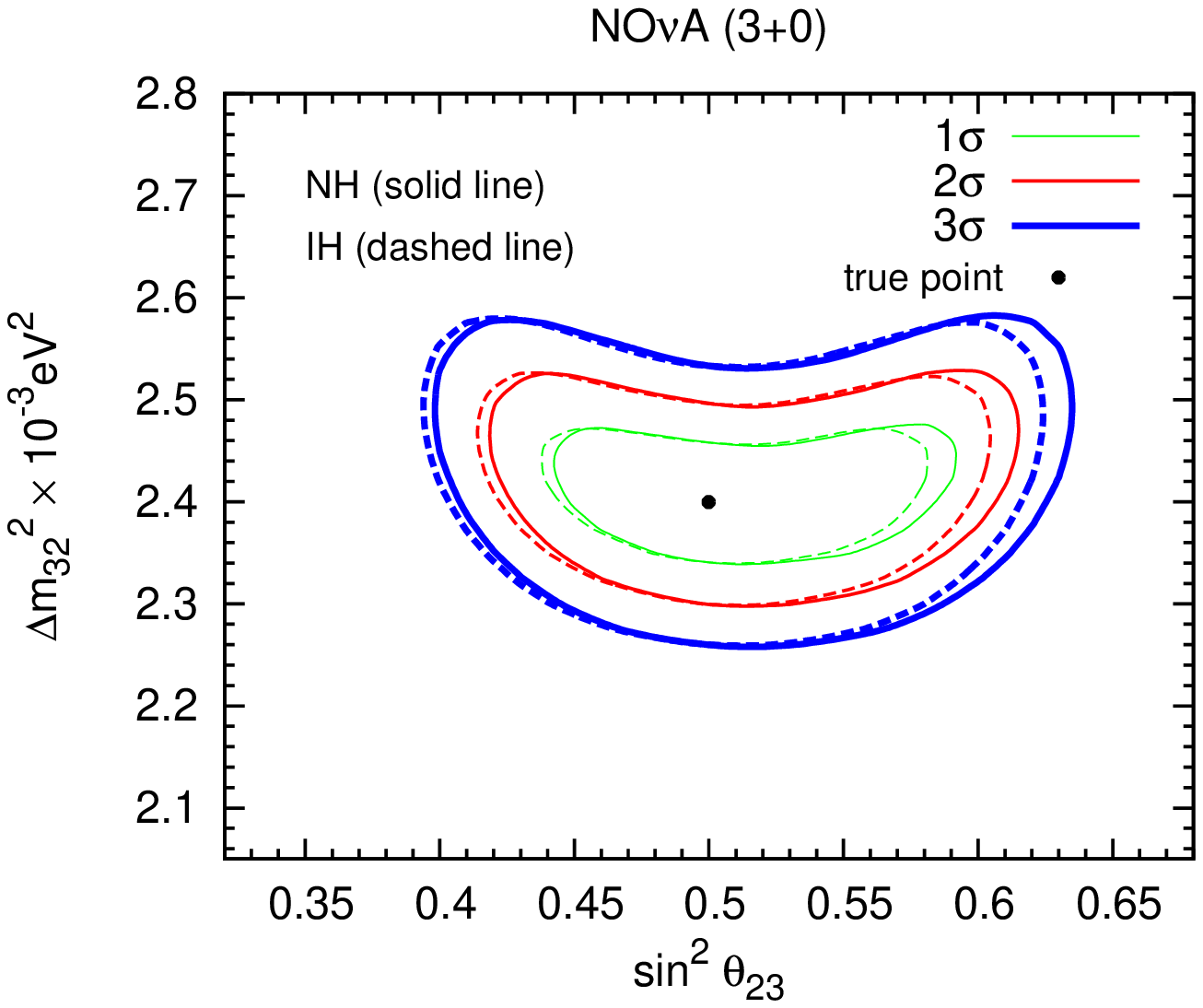}\\
\includegraphics[width=7cm,height=7cm]{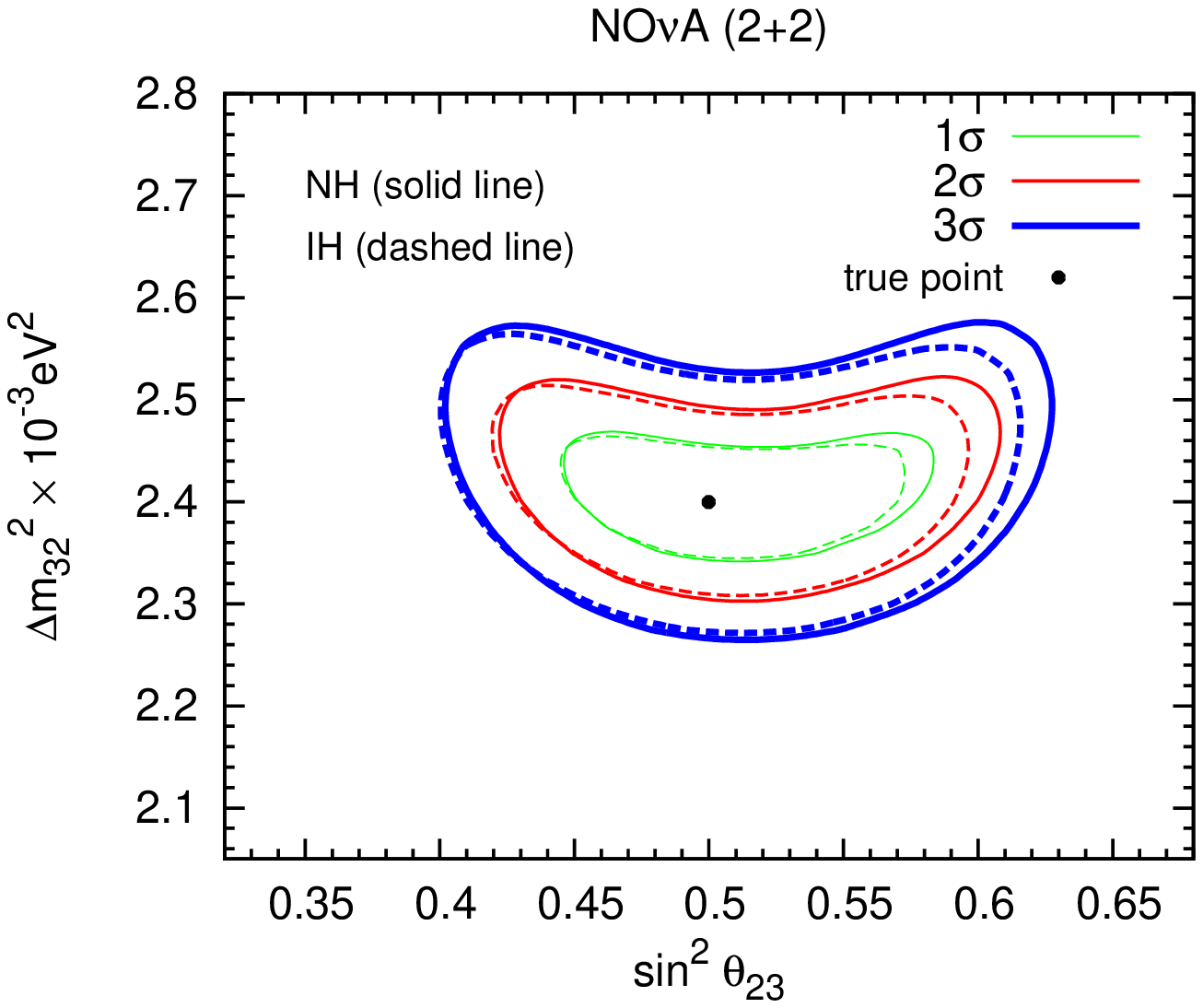}
\includegraphics[width=7cm,height=7cm]{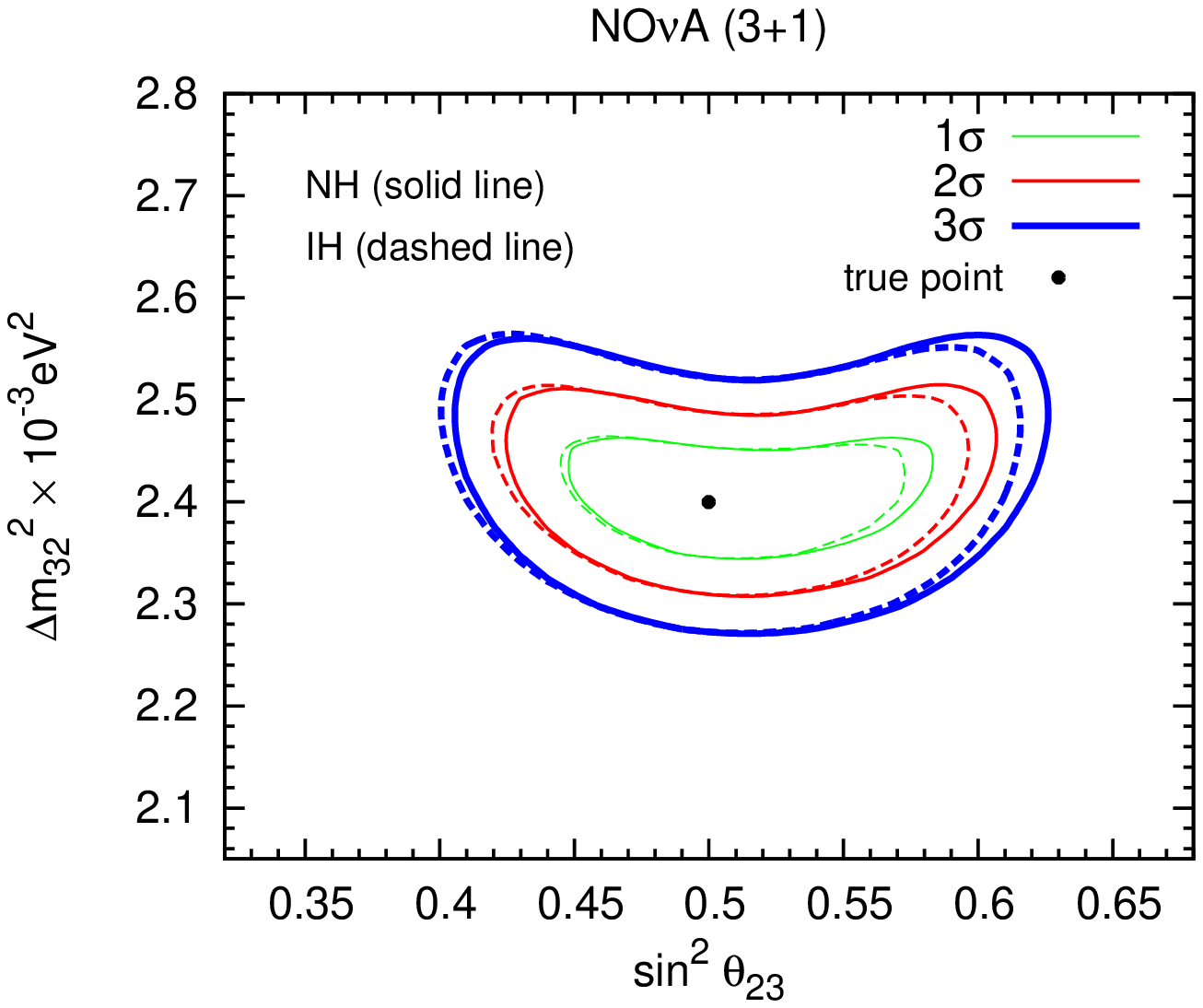}\\
\includegraphics[width=7cm,height=7cm]{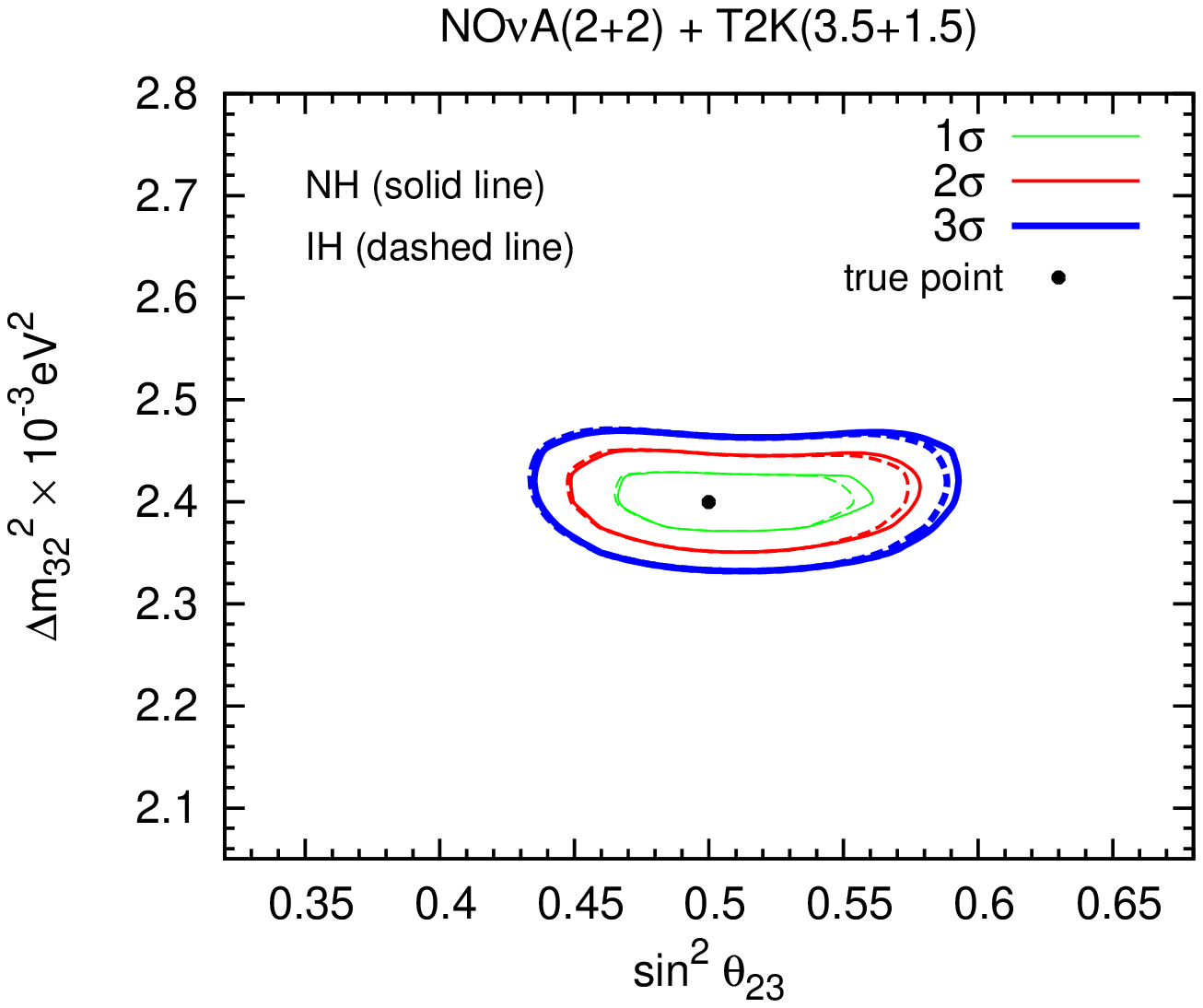}\\
\caption{The $\sin^2\theta_{23} - \Delta m^2_{32}$ contour plots with true $\sin^2\theta_{23} = 0.5$.}
\label{contourt23m23}
\end{figure}

\begin{figure}[!htb]
\includegraphics[width=4cm,height=4cm]{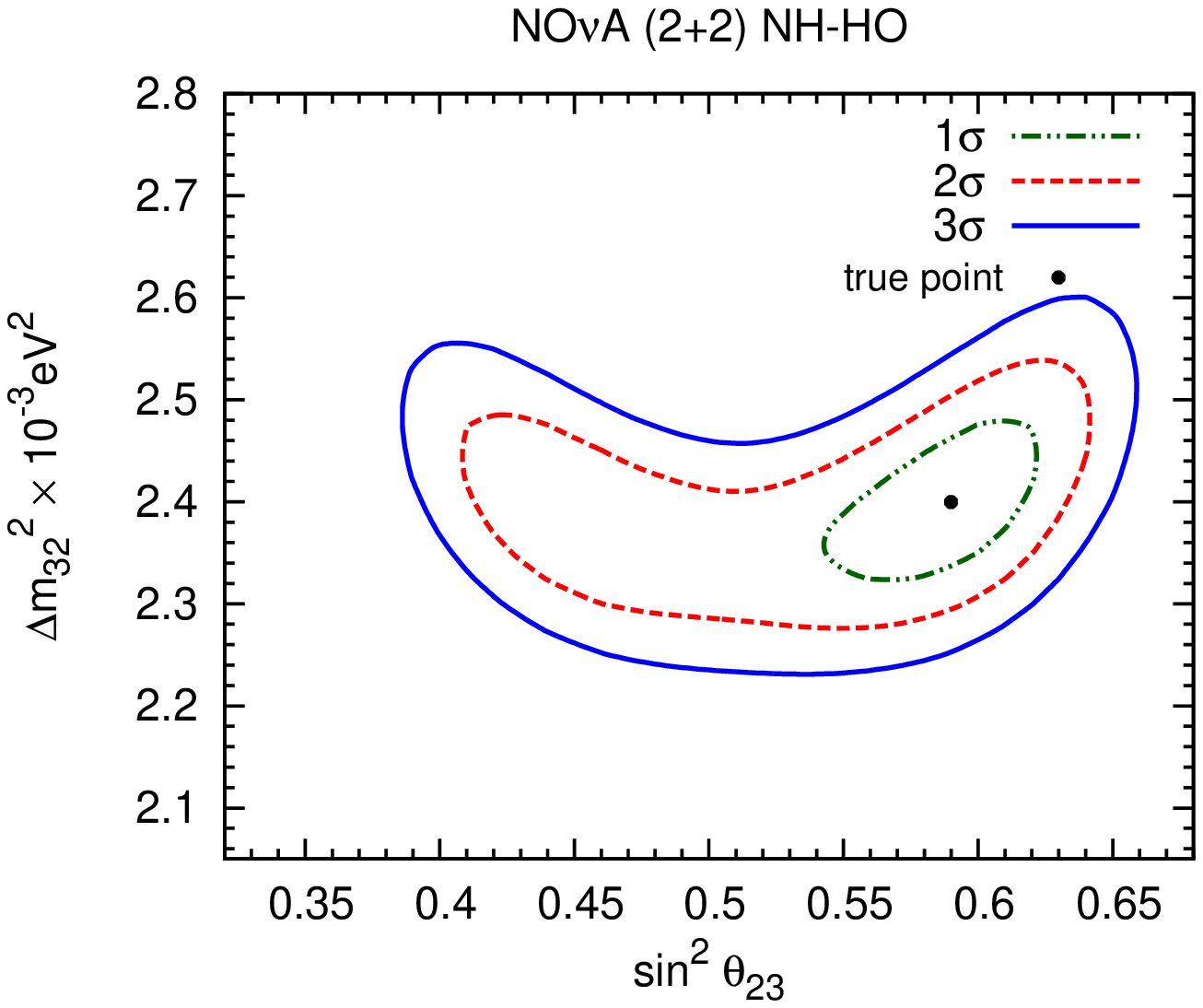}
\includegraphics[width=4cm,height=4cm]{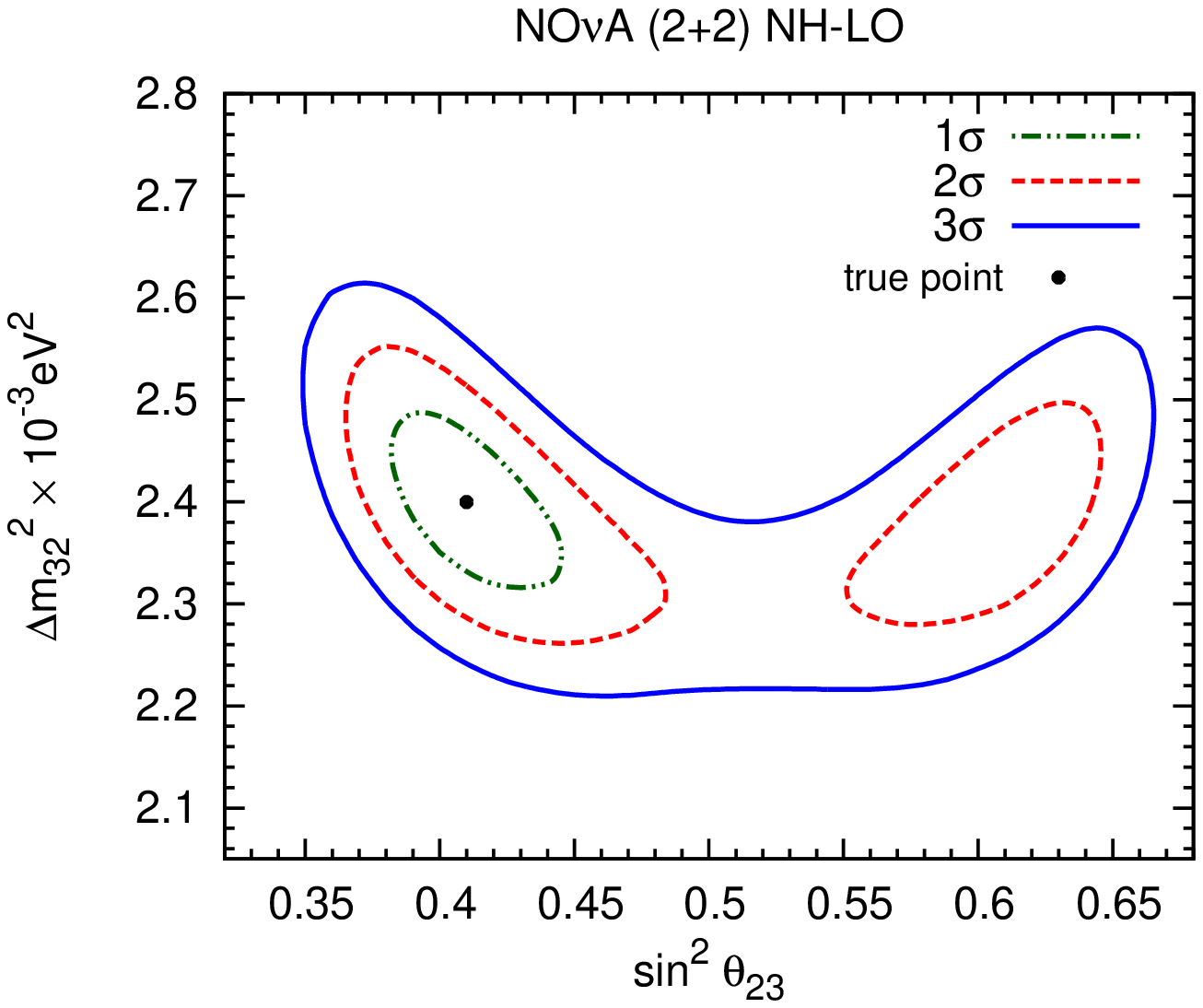}
\includegraphics[width=4cm,height=4cm]{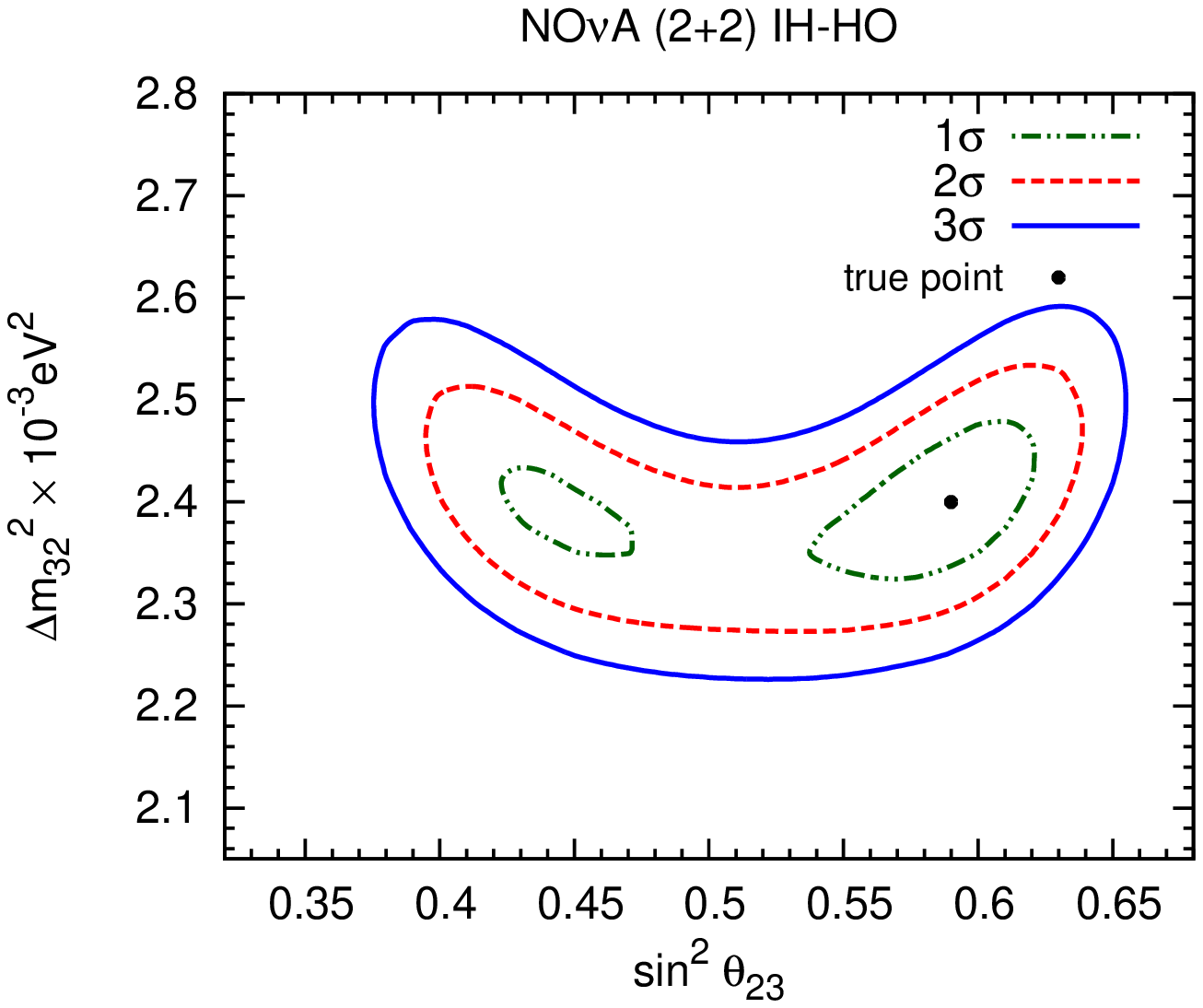}
\includegraphics[width=4cm,height=4cm]{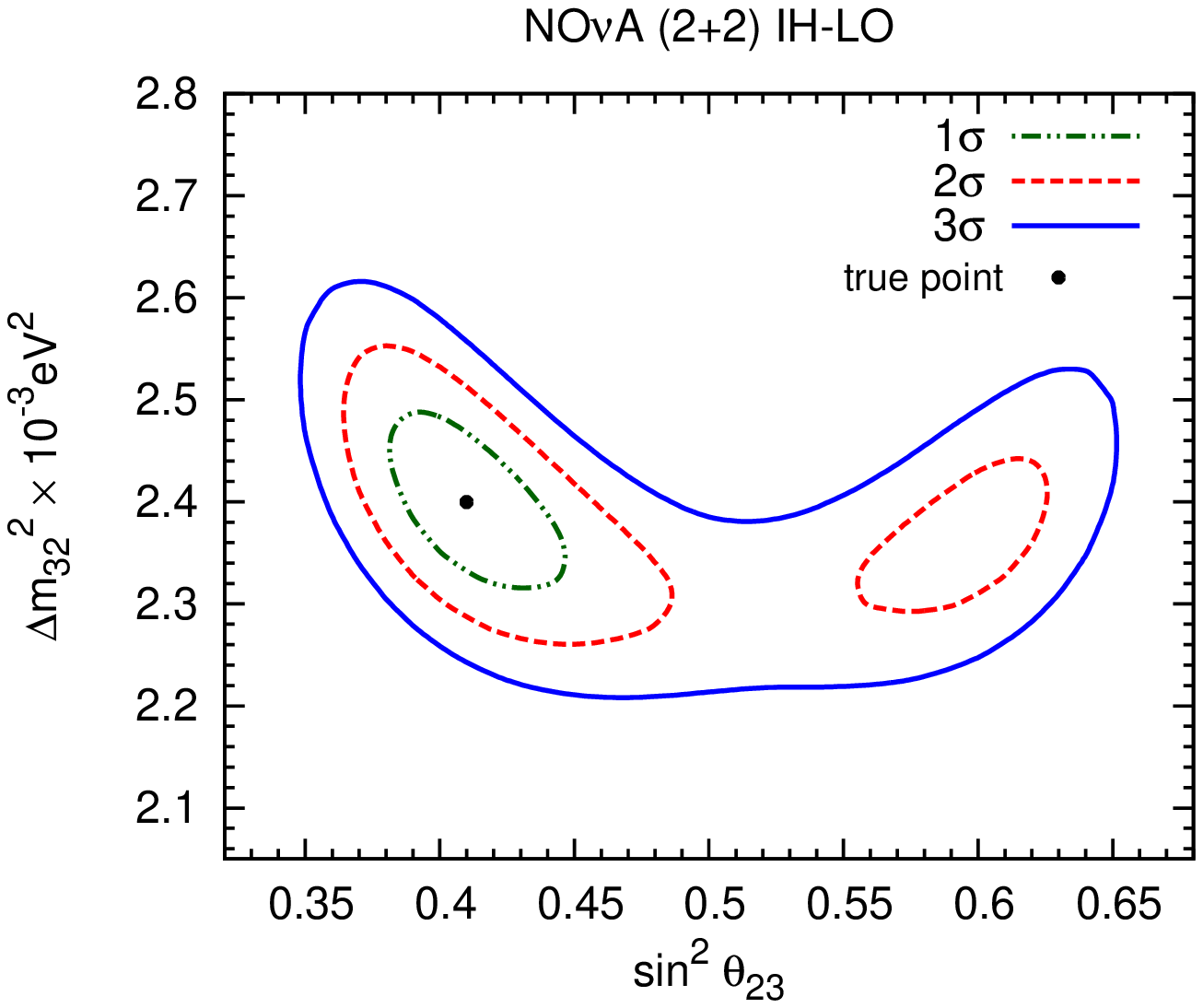}\\
\includegraphics[width=4cm,height=4cm]{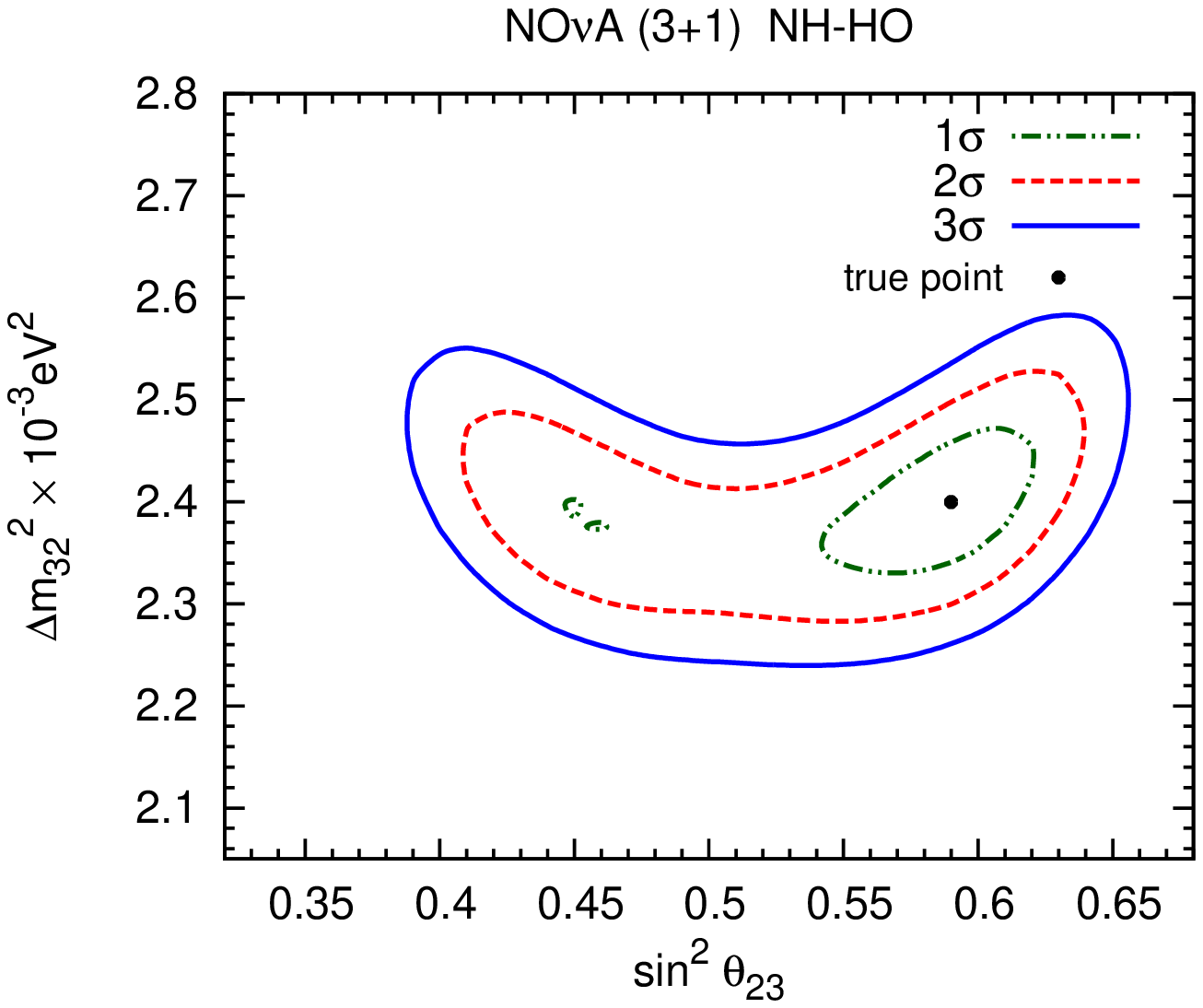}
\includegraphics[width=4cm,height=4cm]{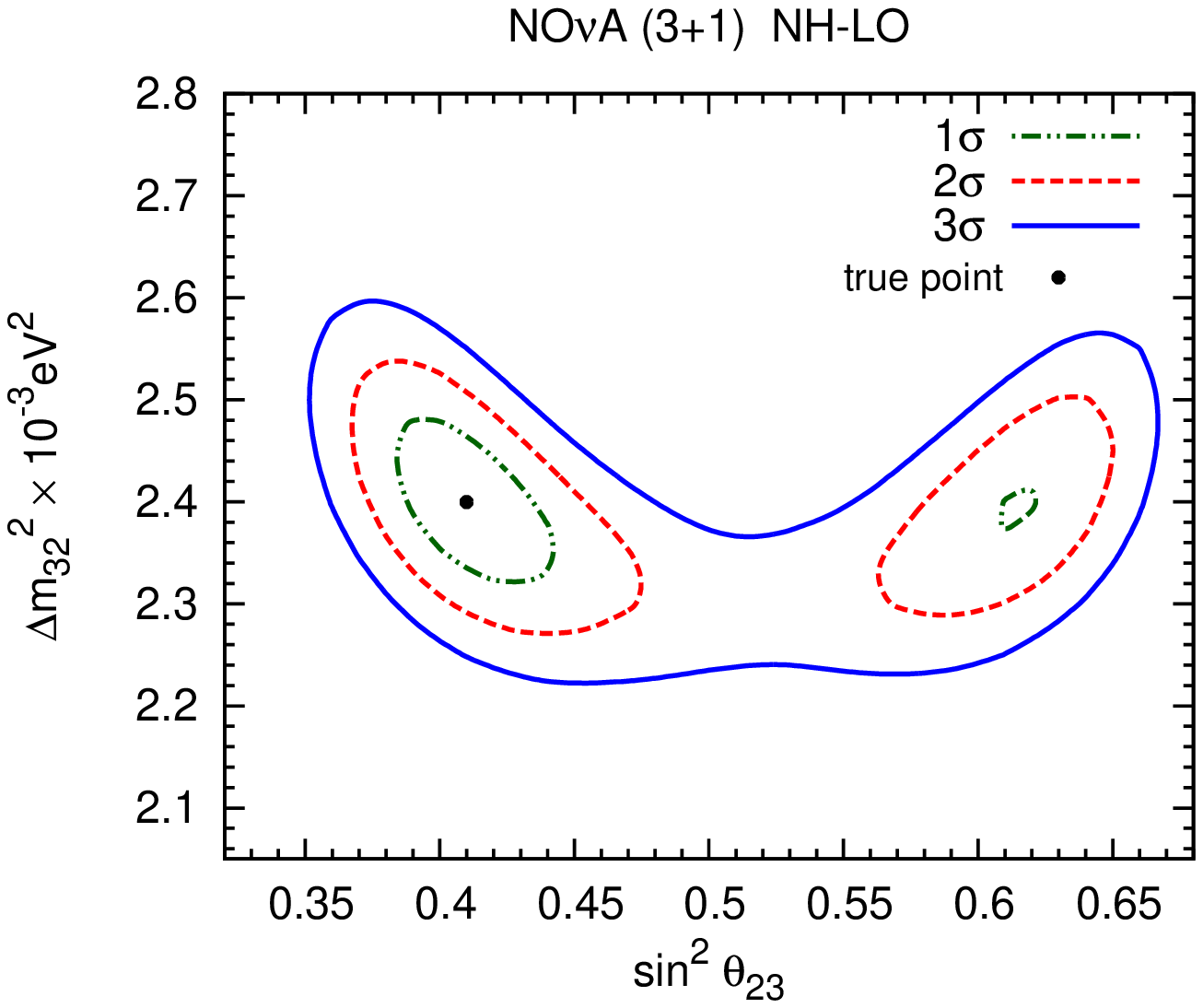}
\includegraphics[width=4cm,height=4cm]{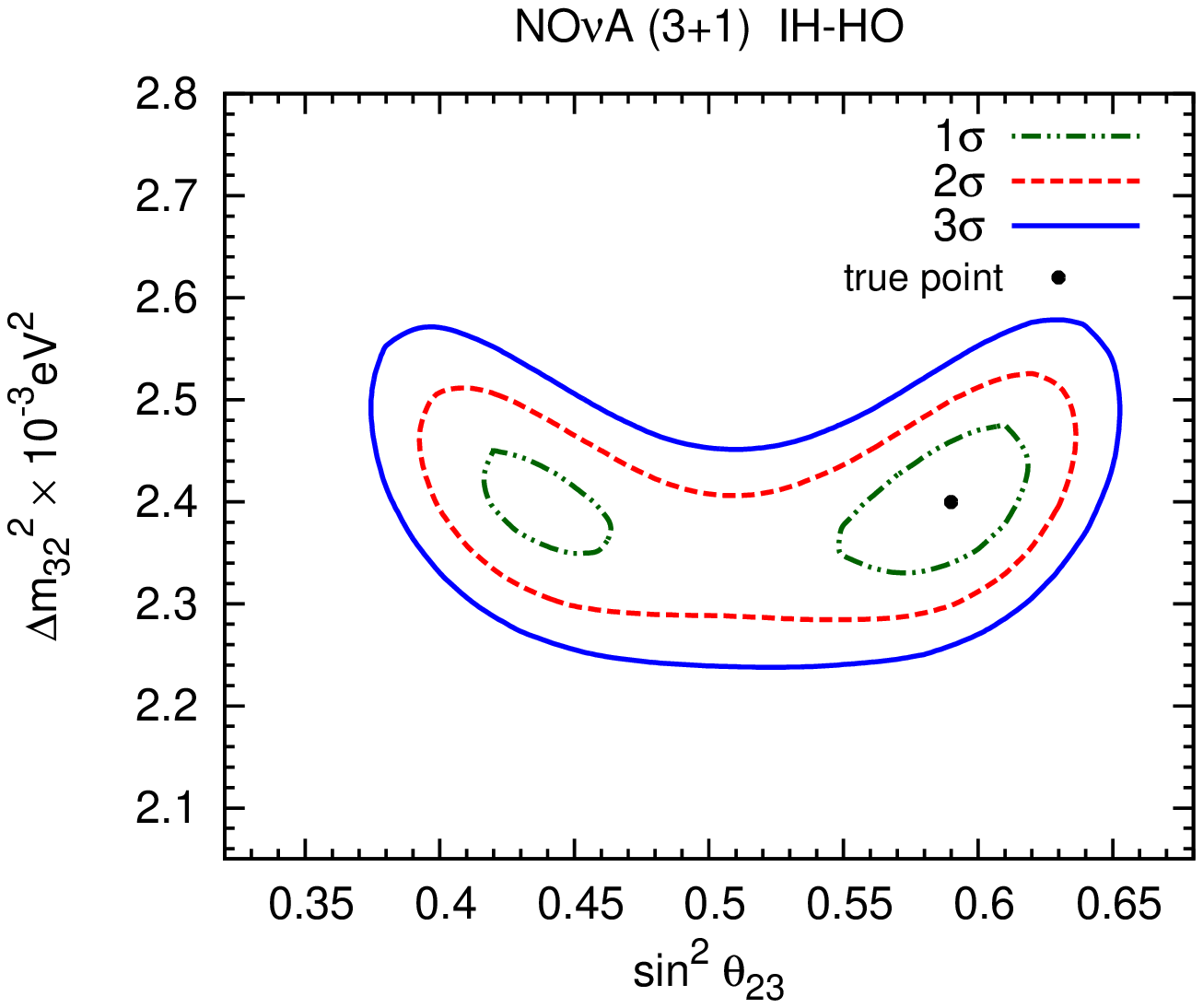}
\includegraphics[width=4cm,height=4cm]{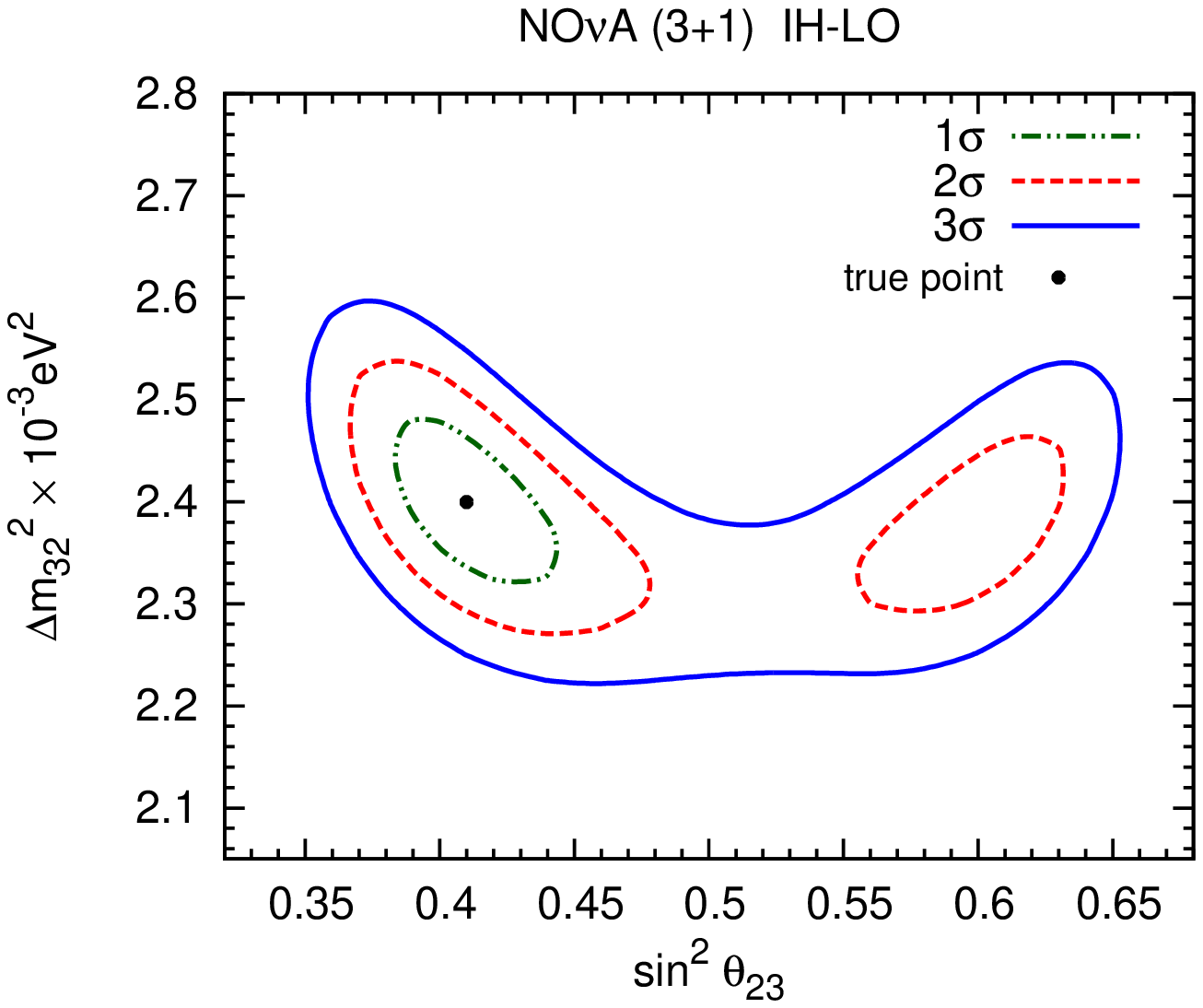}\\
\caption{The $1\sigma$ (green), $2\sigma$ (red), and $3\sigma$ (blue) C.L. regions for  $\sin^2\theta_{23}$ vs. $\Delta m^2_{32}$ with true $\sin^2\theta_{23} =$ 0.41 (0.59) for LO (HO) and true $\delta_{CP}$= 0.}
\label{t23dm23=0}
\end{figure}
In this section, we focus on the   $\theta_{23}$ and $\Delta m^2_{32}$ degeneracy. First of all, we would like to see how does the hierarchy 
ambiguity affect  $\sin^2\theta_{23}$ - $\Delta m^2_{32}$ parameter space. Therefore, we simulate the  true events  for maximal value of 
$\sin^2\theta_{23}$ ($\sin^2\theta_{23} = 0.5$) and  test events for allowed values of  $\sin^2\theta_{23}$ ([0.32:0.68]) and 
$\Delta m^2_{32}$  ($[2.05:2.75]\times 10^{-3}~{\rm eV}^2$ ). We obtain the $\chi^2$ by comparing true events and test events. 
We also do marginalization for both $\sin^22\theta_{13}$ and $\delta_{CP}$ and add a 
prior on  $\sin^22\theta_{13}$. In Fig. \ref{contourt23m23}, the obtained  $\chi^2$ is plotted as a function of  $\sin^2\theta_{23}$ and $\Delta m^2_{32}$. 
From the plots, we can see that there is  small difference in the allowed parameter space for NH and IH. However, 
there is no difference in the allowed parameter space 
for (2+2) and (3+1) years of NO$\nu$A running as far as the determination of $\Delta m_{32}^2$ is concerned. 
We also get  similar results   when we compare  the parameter space for both NO$\nu$A (2+2) 
and NO$\nu$A (3+1), and as expected such parameter spaces are significantly reduced when compared with NO$\nu$A (2+1) and NO$\nu$A (3+0). 
It should also be noted from the figure that the parameter 
space is substantially reduced for a combined analysis of T2K and NO$\nu$A. Therefore, if we combine the (2+2) years of NO$\nu$A results with (3.5+1.5) 
T2K results, the significance of the atmospheric mass square determination will improve significantly.

We also obtain  $\sin^2\theta_{23}$ - $\Delta m^2_{32}$ parameter space for non-maximal mixing of atmospheric mixing angle. 
We  consider deviation from maximal mixing with $\sin^2\theta_{23}$ = 0.41 (0.59) for Lower Octant (Higher Octant). Fig \ref{t23dm23=0} 
shows the $\sin^2\theta_{23}$ - $\Delta m^2_{32}$ parameter space for NH-HO, NH-LO, IH-HO, IH-LO combinations. It is clear from the figures 
that, in this case also there  is no significant difference between the allowed parameter space for (2+2) and (3+1) years of  NO$\nu$A run period.  
Therefore, the expected results  on $\theta_{23}$ and $\Delta m^2_{32}$ degeneracy discrimination 
would not be deteriorated if NO$\nu$A switches to antineutrino mode after completion of 2 years of neutrino run.

\subsection{Correlation between  $\delta_{CP}$ and $\sin^2\theta_{23}$  }

\begin{figure}[!htb]
\includegraphics[width=4cm,height=4cm]{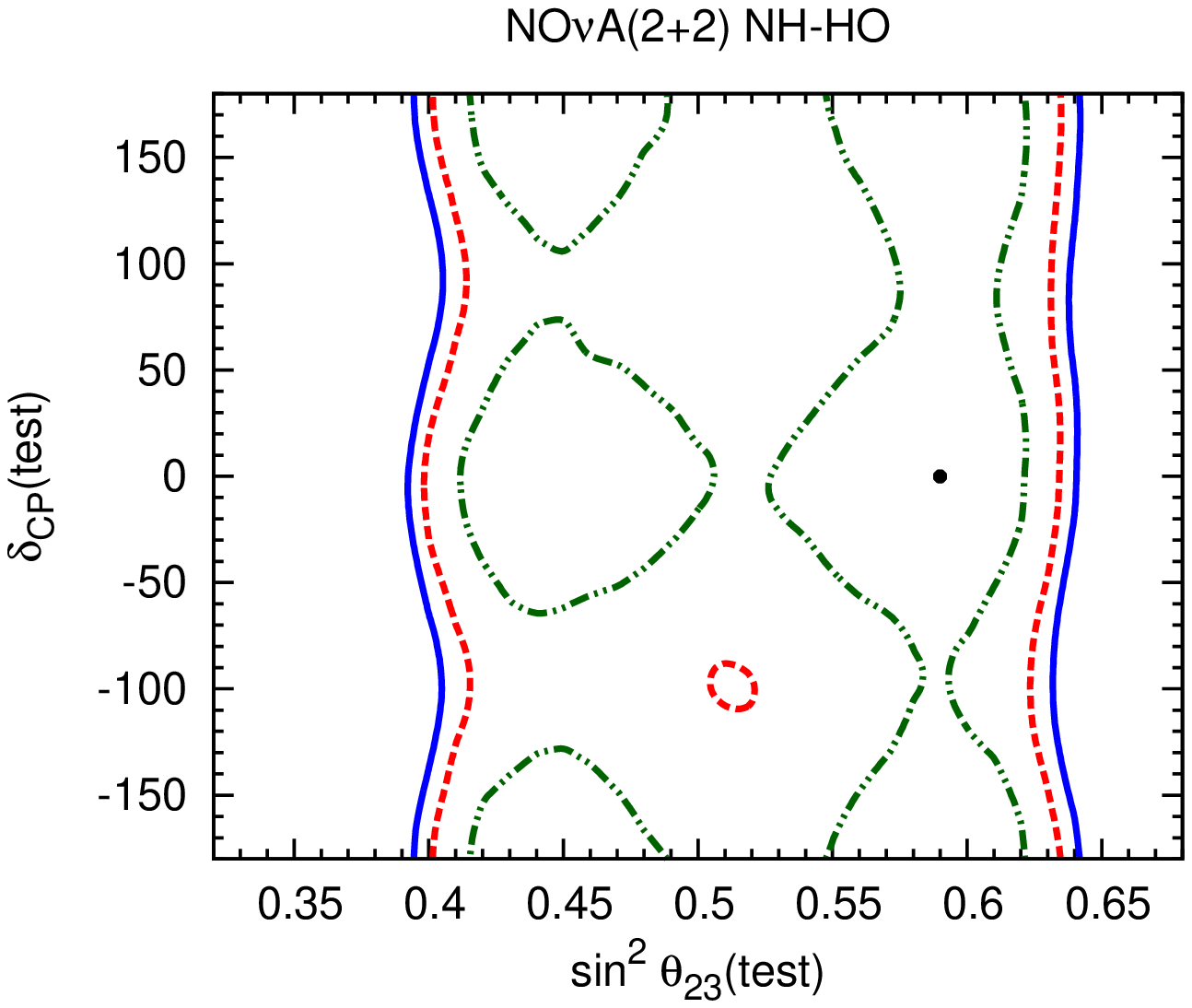}
\includegraphics[width=4cm,height=4cm]{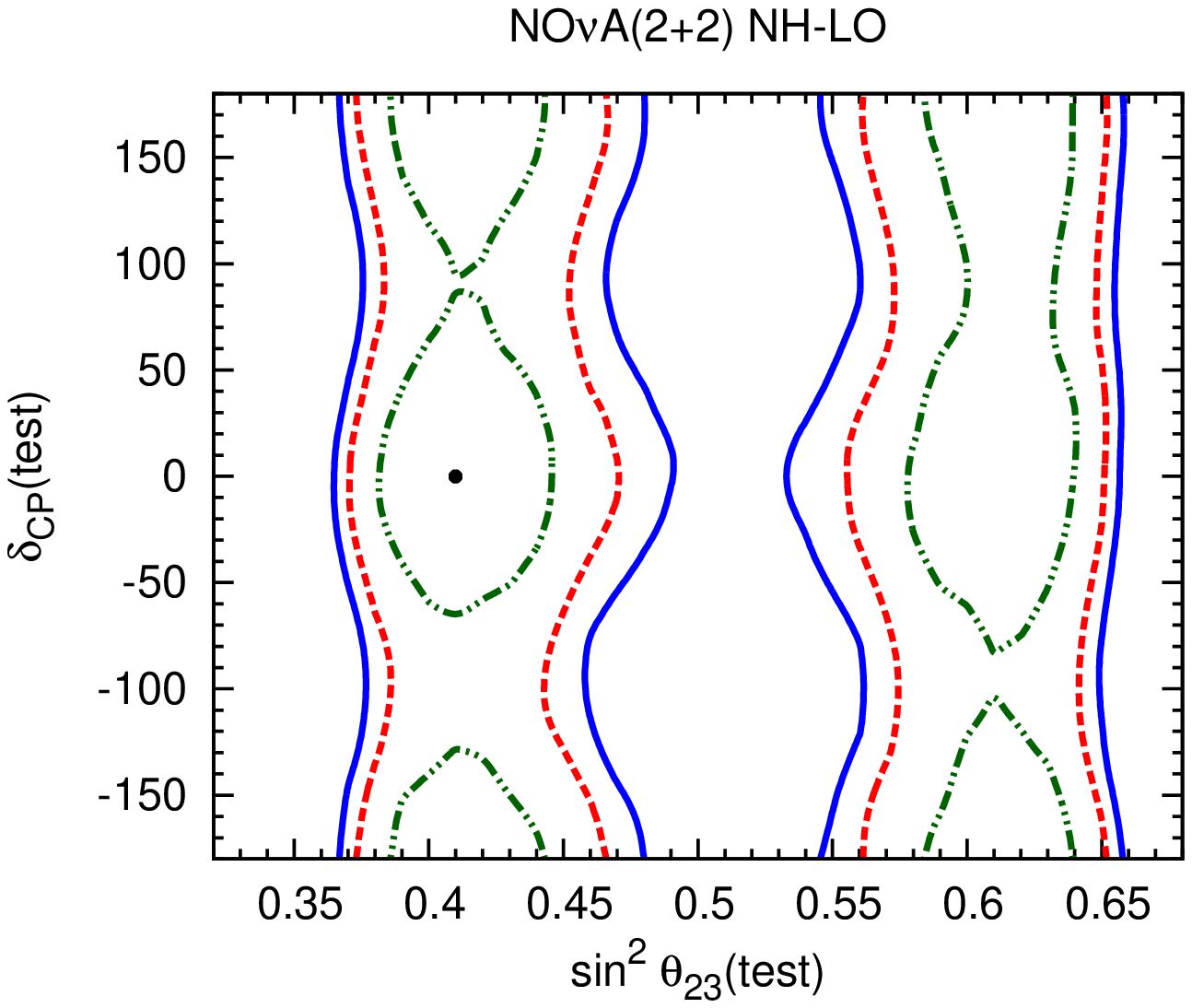}
\includegraphics[width=4cm,height=4cm]{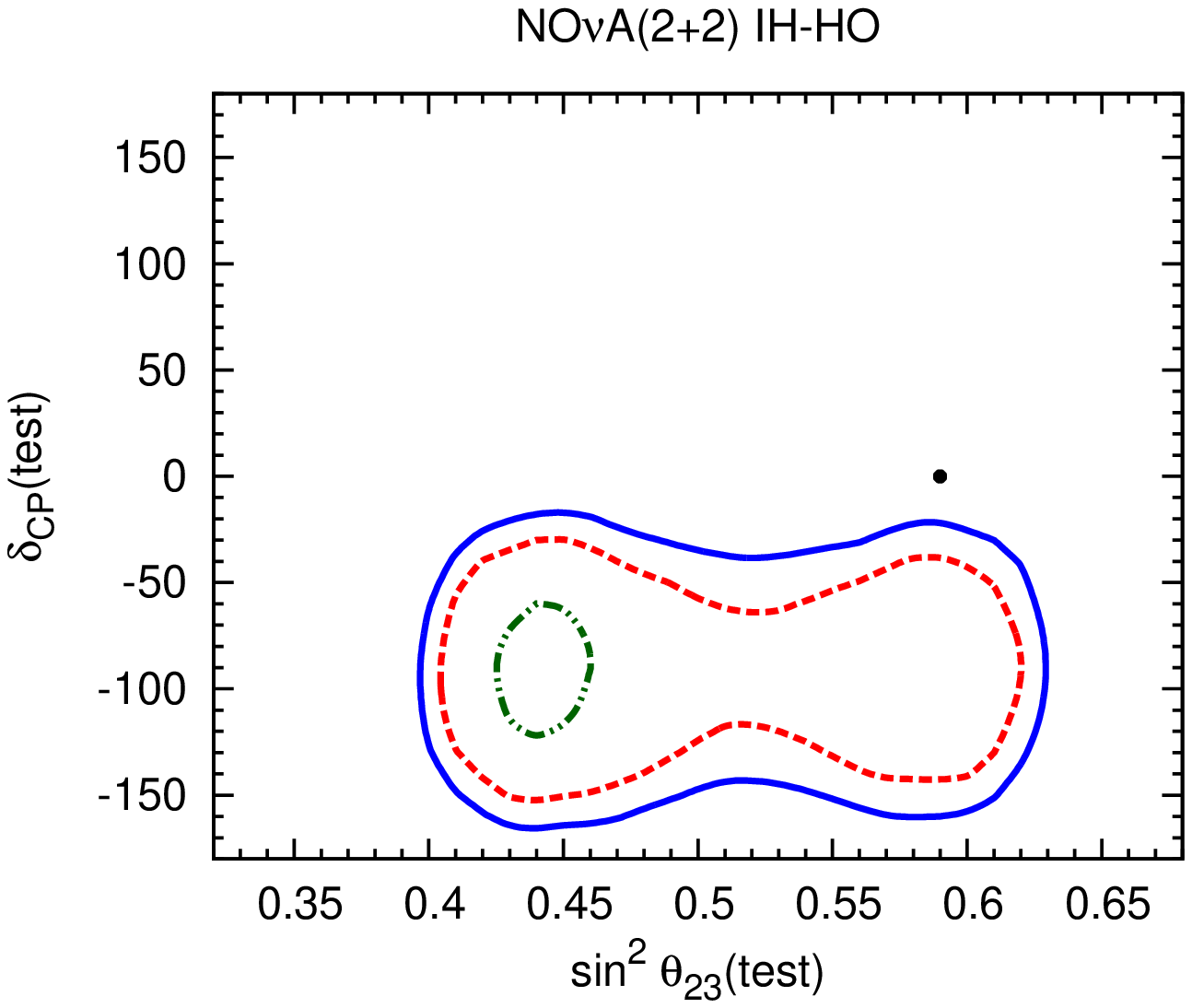}
\includegraphics[width=4cm,height=4cm]{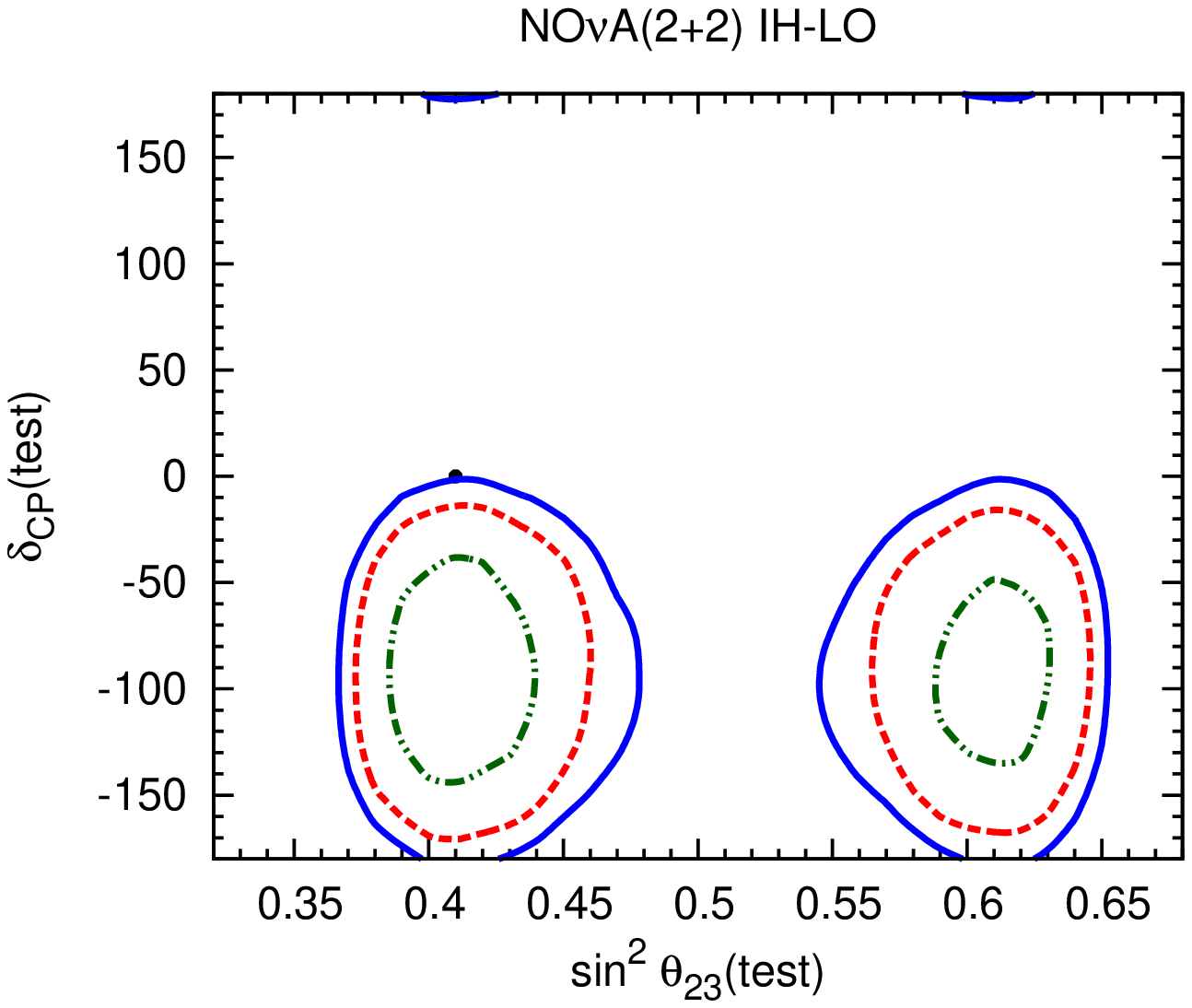}\\
\includegraphics[width=4cm,height=4cm]{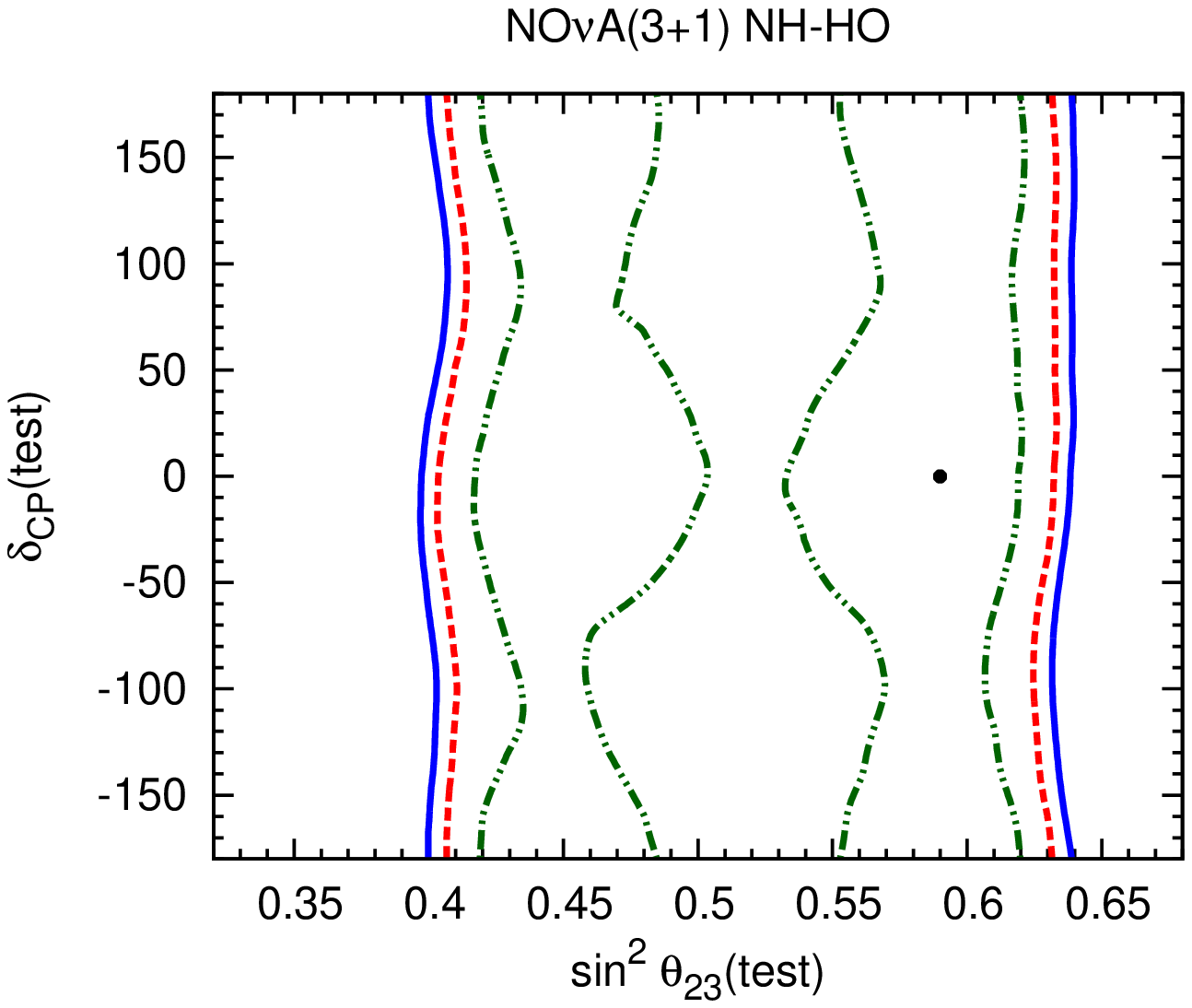}
\includegraphics[width=4cm,height=4cm]{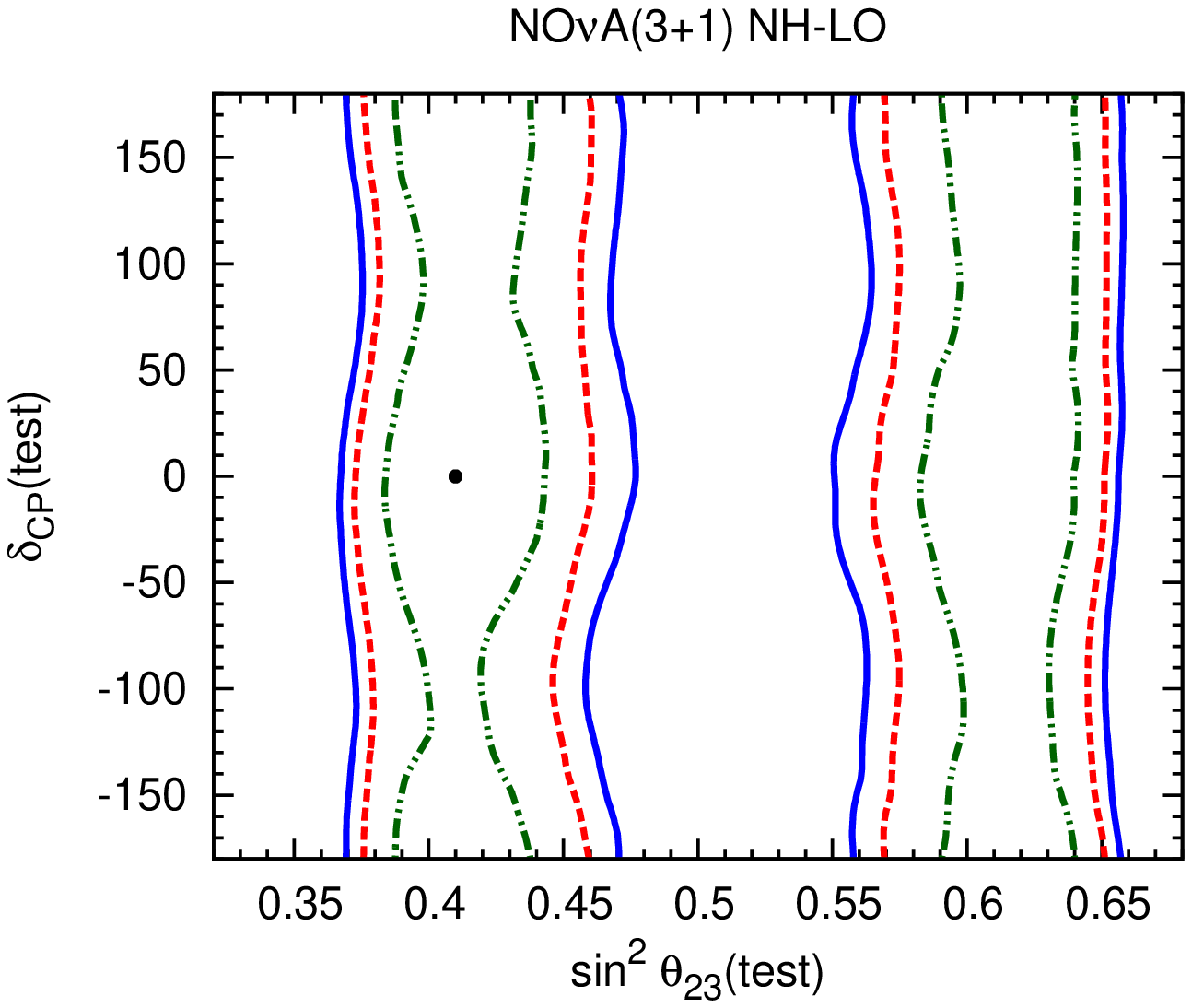}
\includegraphics[width=4cm,height=4cm]{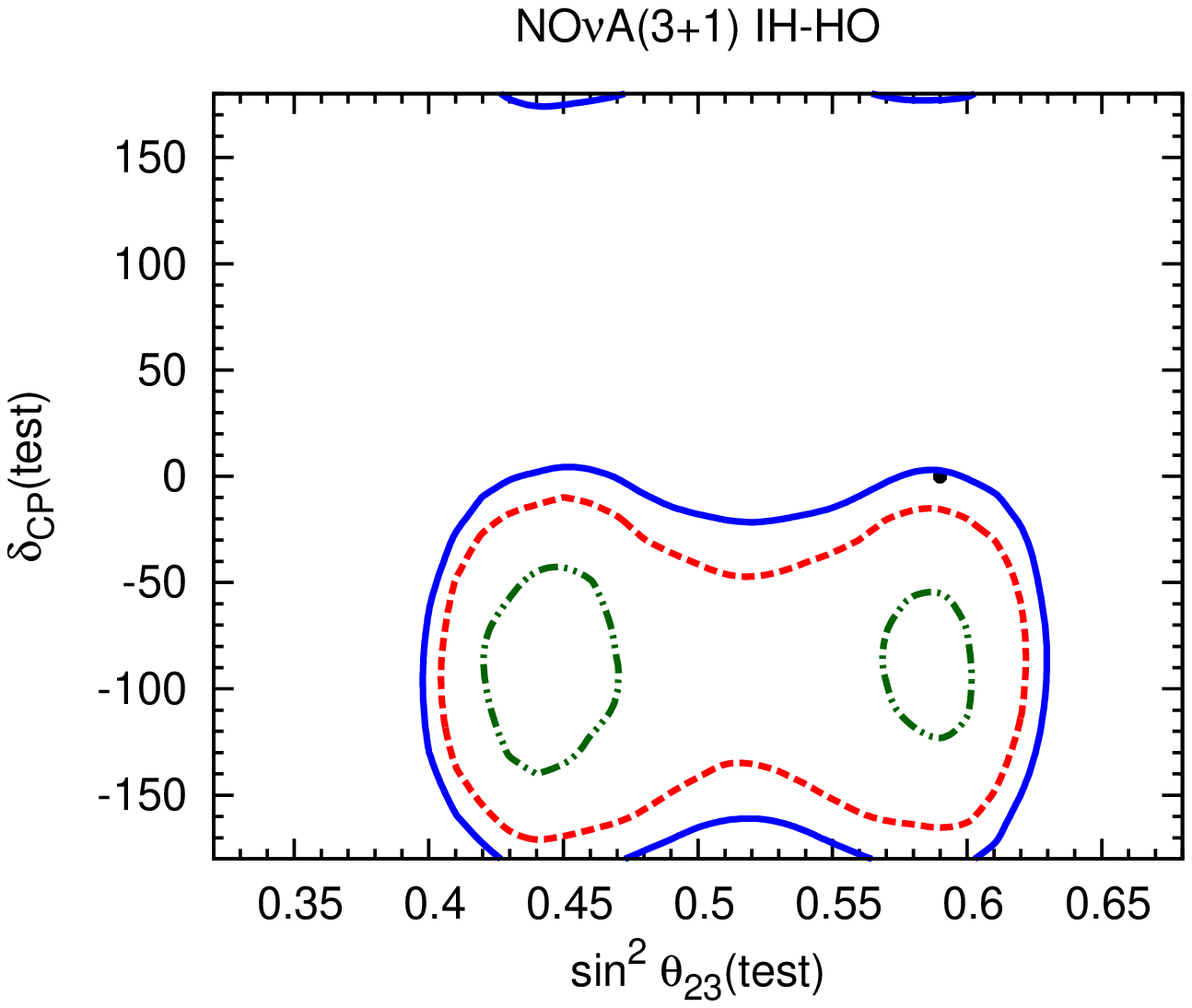}
\includegraphics[width=4cm,height=4cm]{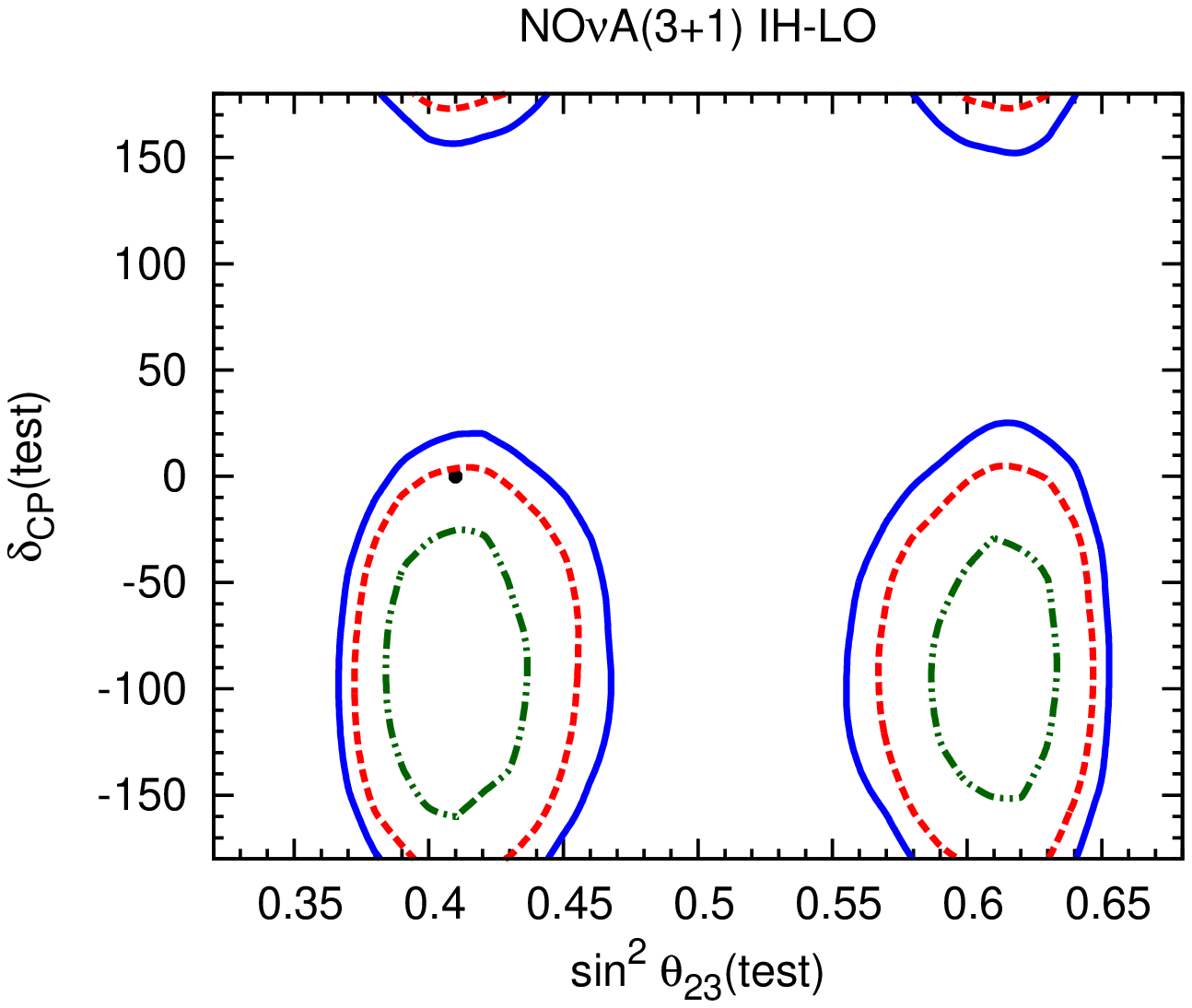}\\
\caption{The $1\sigma$ (green), $2\sigma$ (red), and 90\% (blue) C.L. regions for  $\sin^2\theta_{23}~ {\rm vs.}~ \delta_{CP}$ 
with true $\sin^2\theta_{23} =$ 0.41(0.59) for LO(HO) and true $\delta_{CP}$= 0.}
\label{t23dcp=0}
\end{figure}

\begin{figure}[!htb]
\includegraphics[width=4cm,height=4cm]{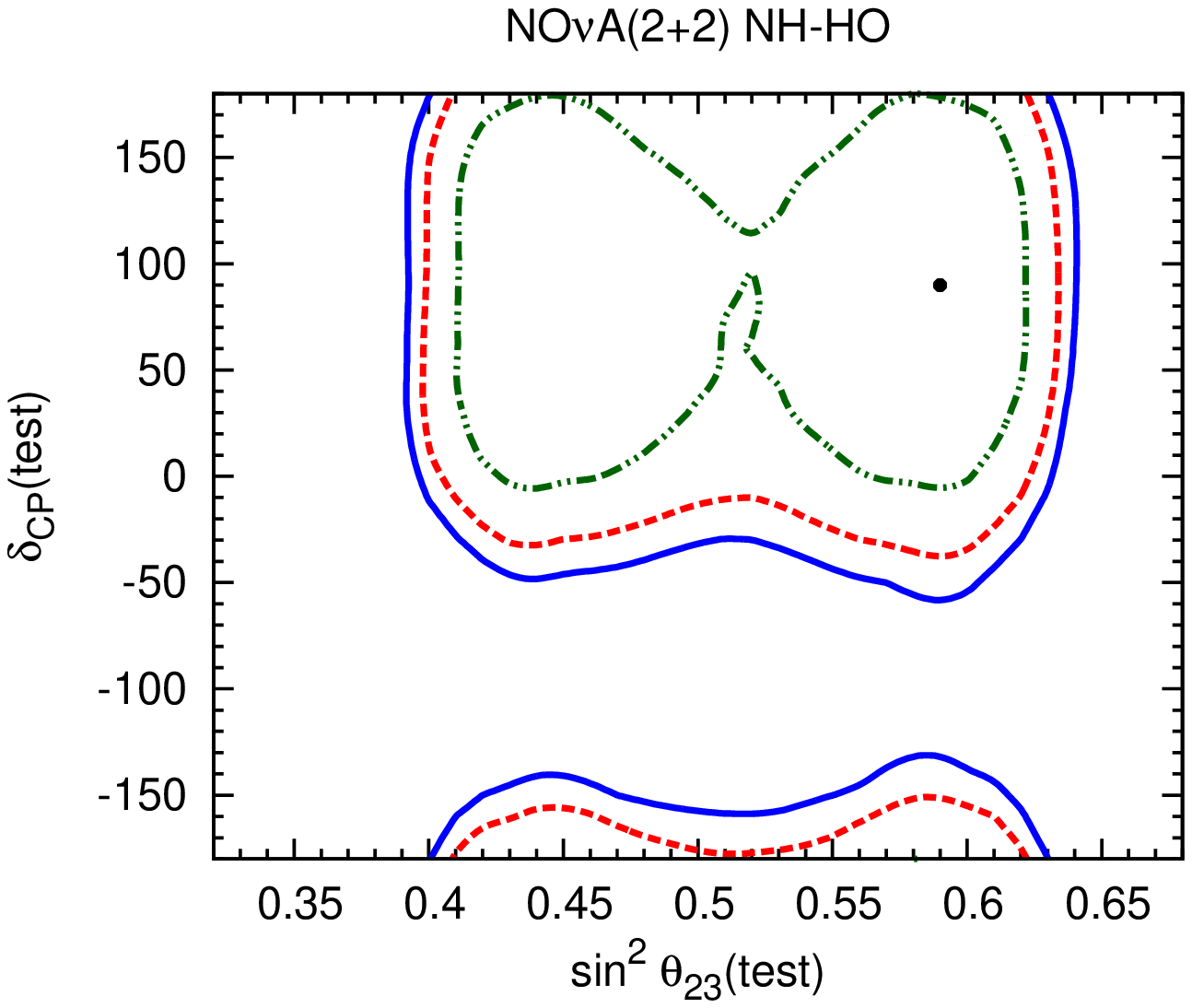}
\includegraphics[width=4cm,height=4cm]{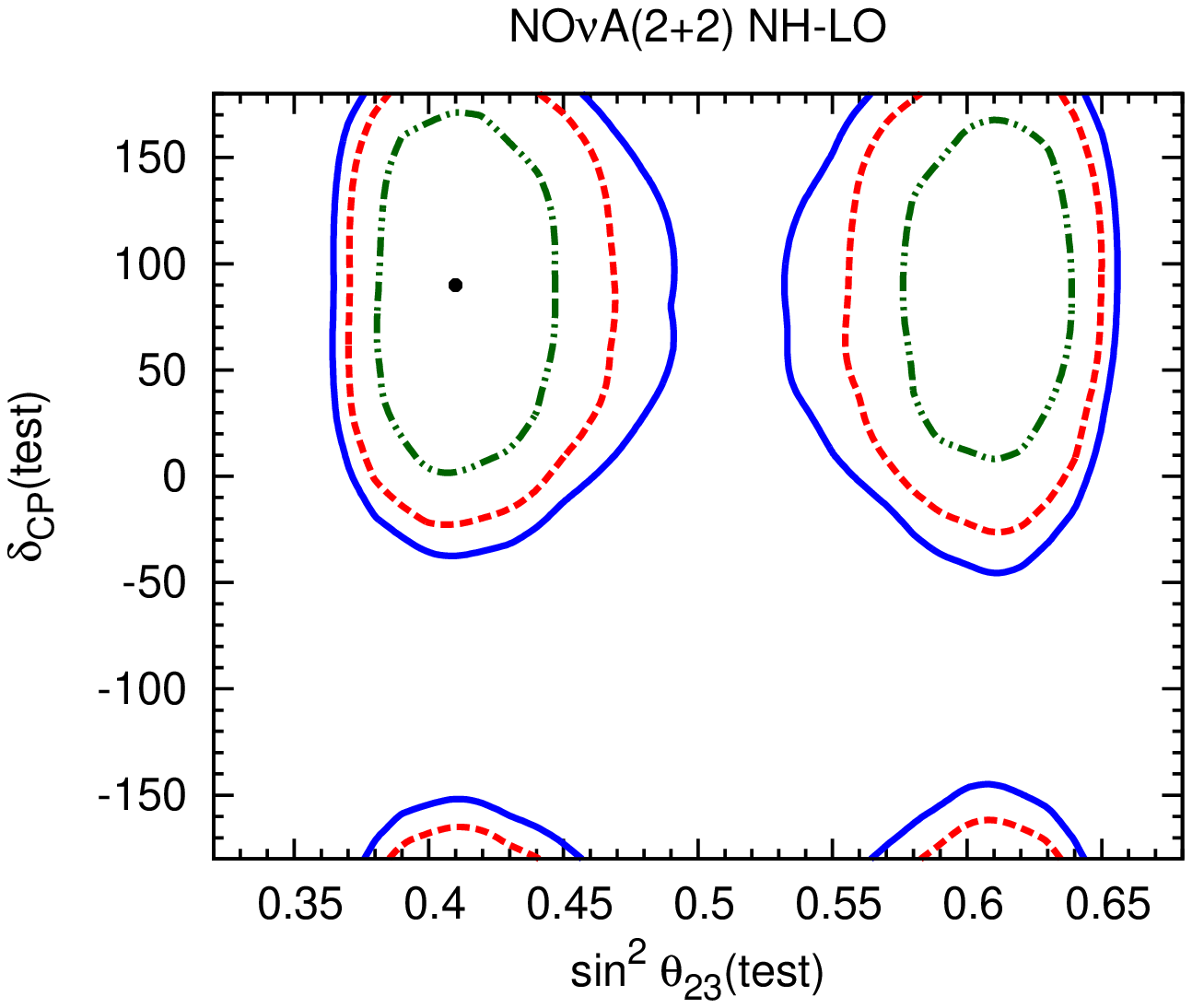}
\includegraphics[width=4cm,height=4cm]{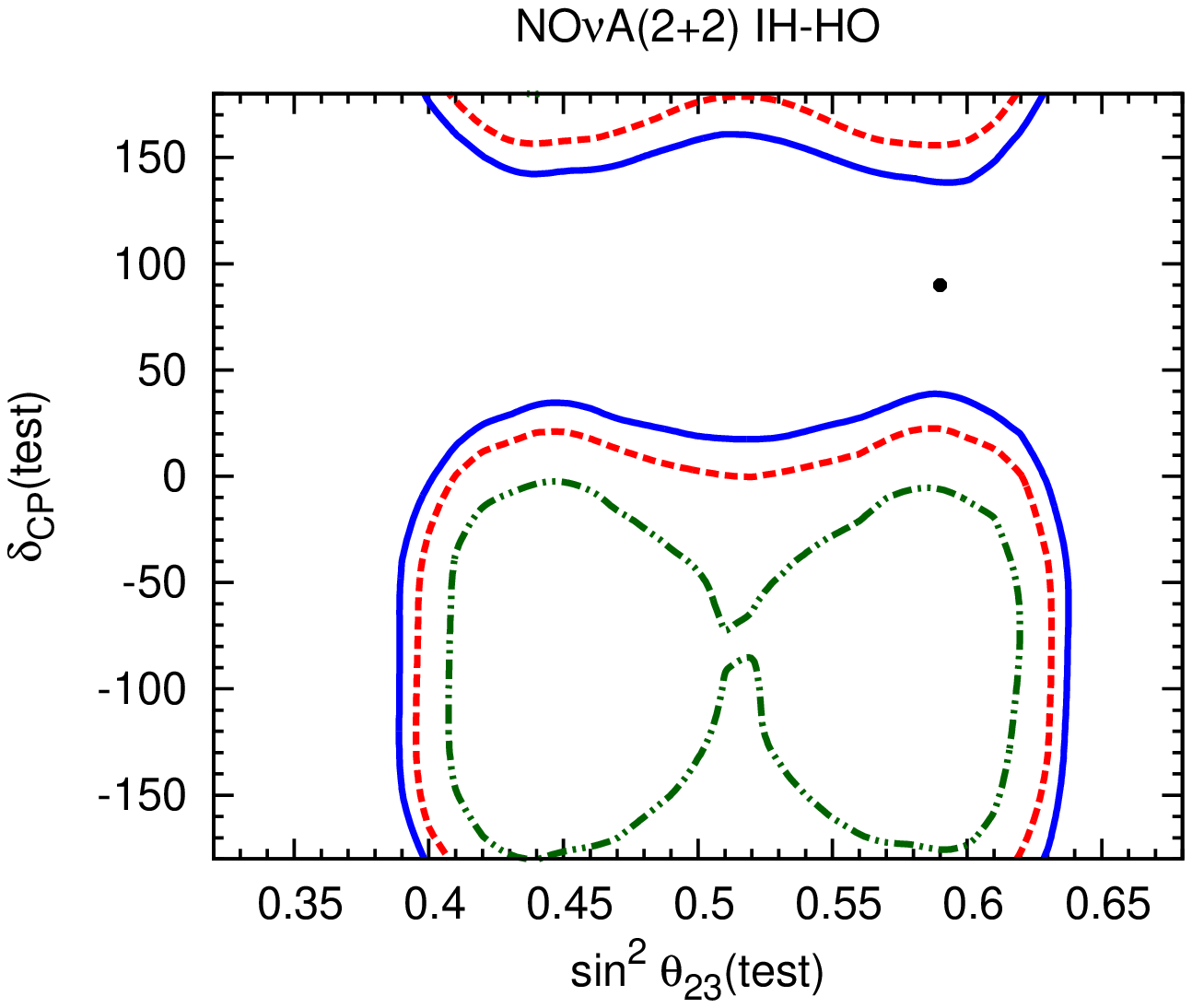}
\includegraphics[width=4cm,height=4cm]{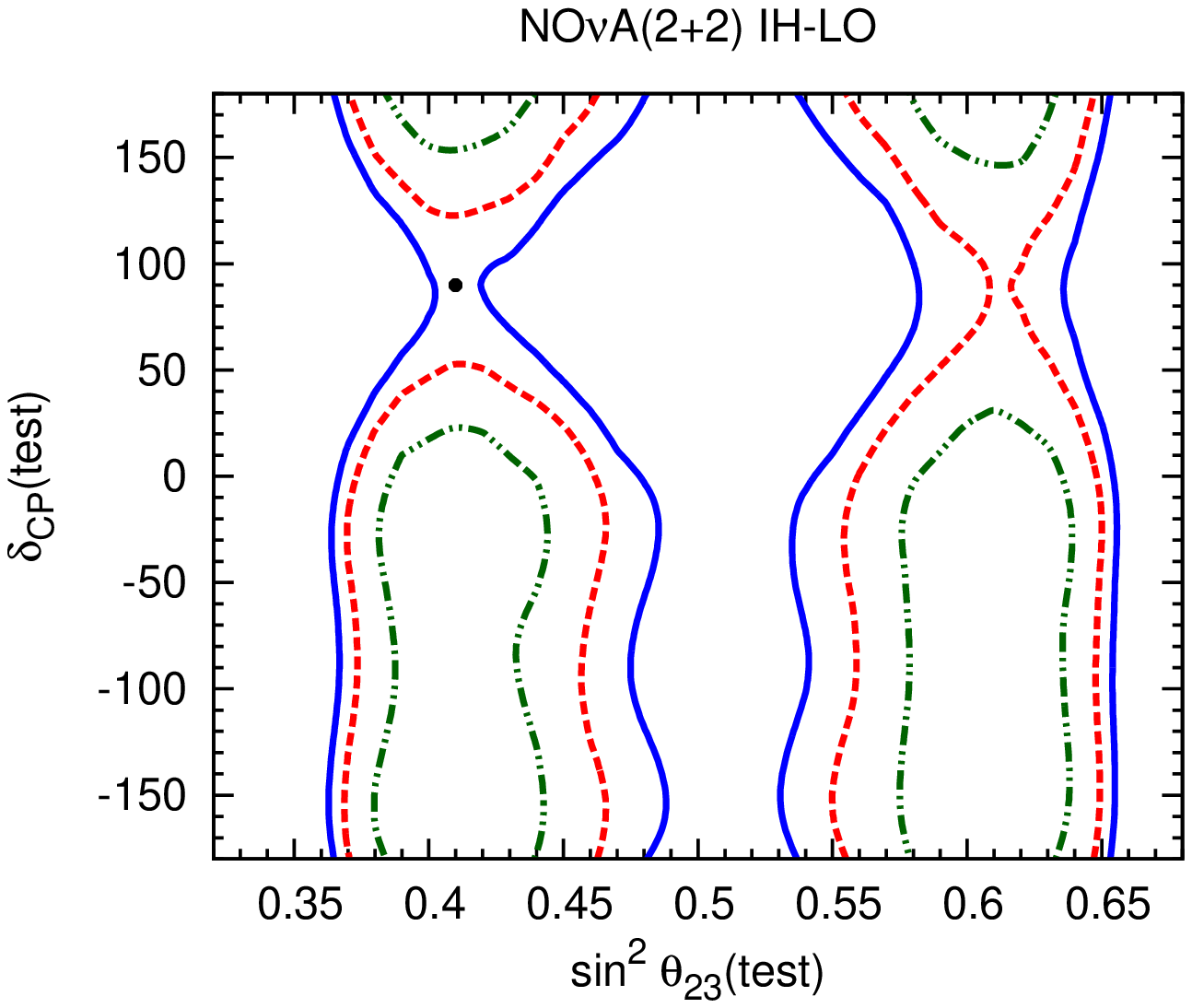}\\
\includegraphics[width=4cm,height=4cm]{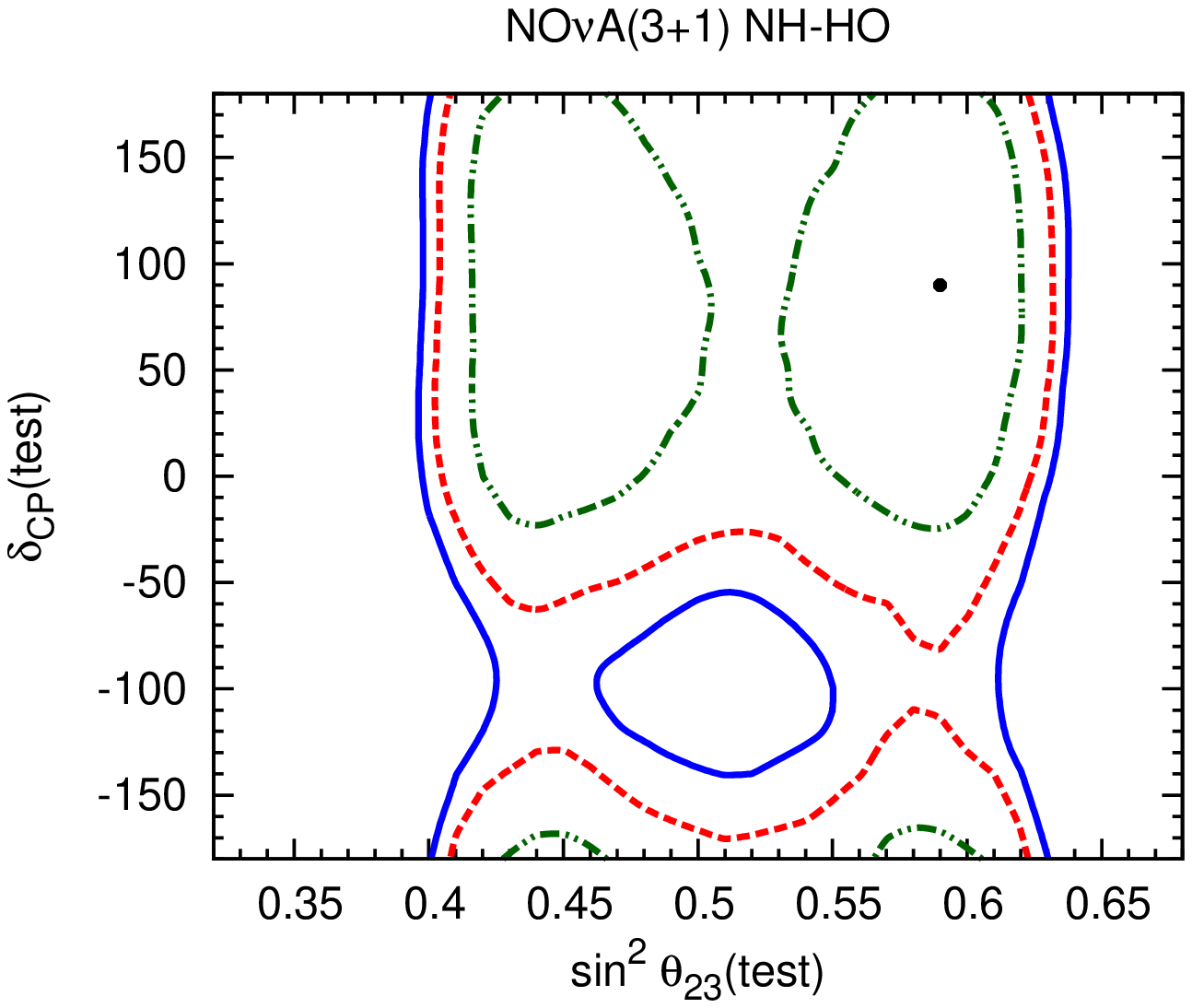}
\includegraphics[width=4cm,height=4cm]{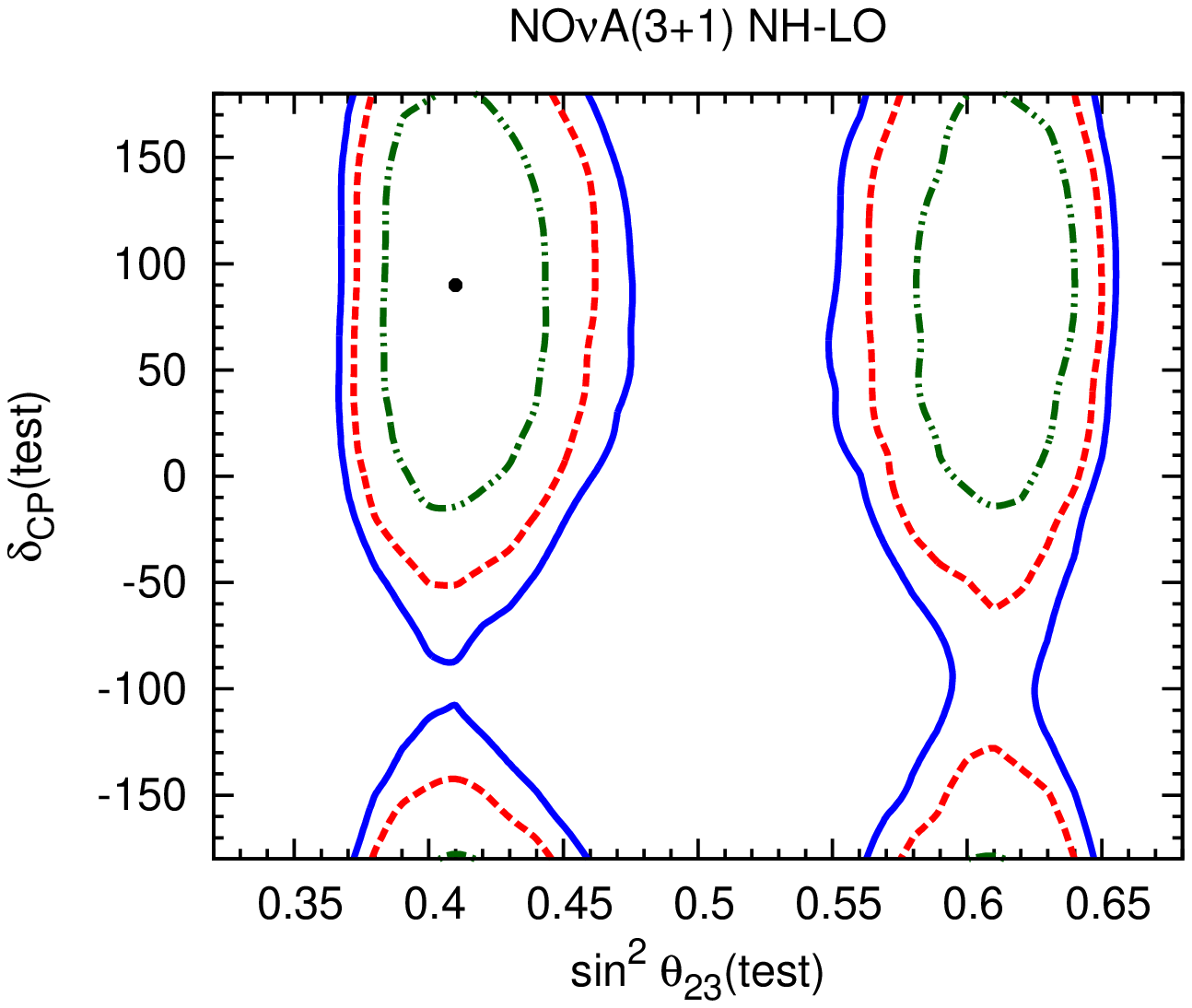}
\includegraphics[width=4cm,height=4cm]{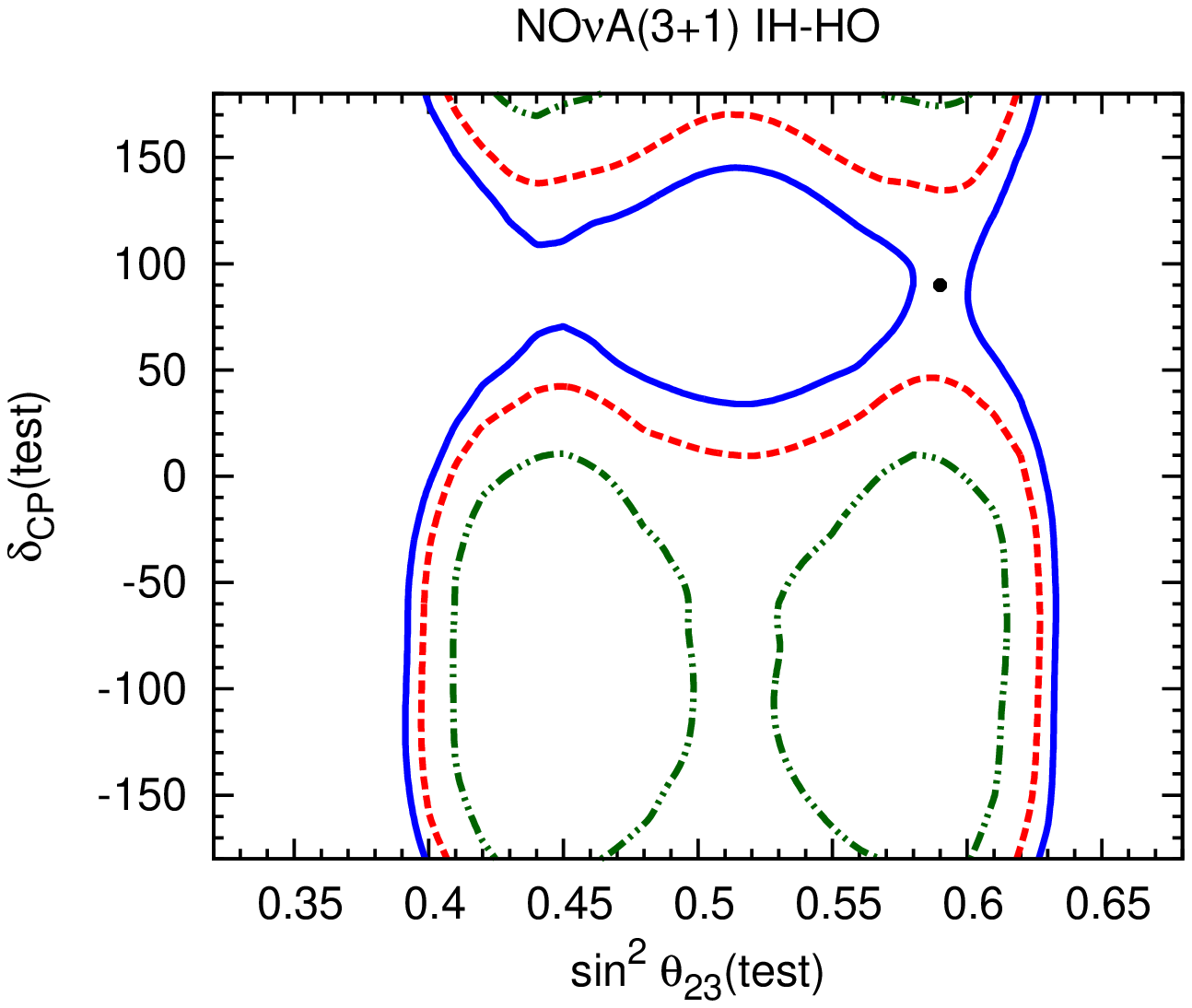}
\includegraphics[width=4cm,height=4cm]{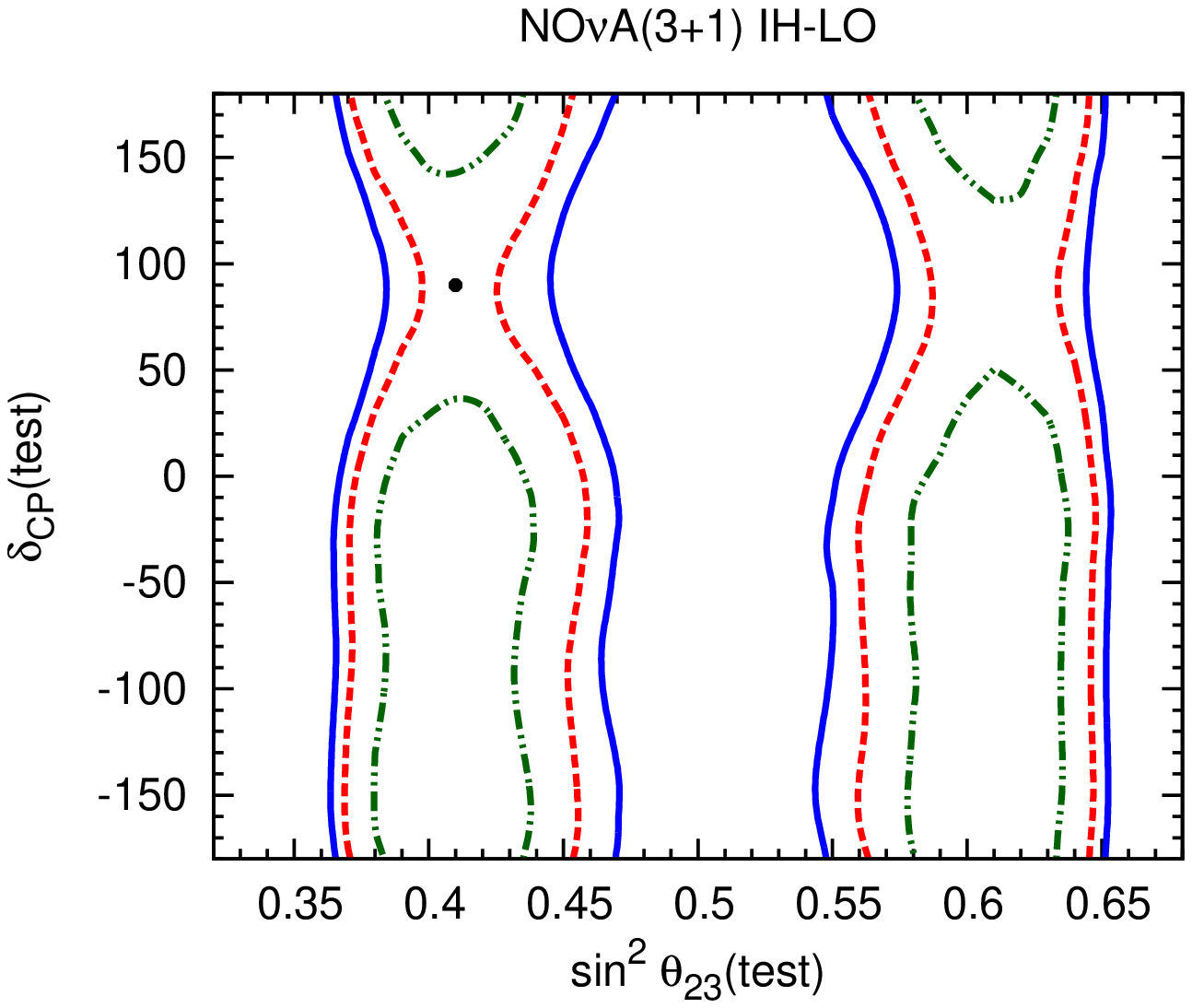}\\
\caption{The $1\sigma$ (green), $2\sigma$ (red), and 90\% (blue) C.L. regions for  $\sin^2\theta_{23}~ {\rm vs.}~ \delta_{CP}$ with true 
$\sin^2\theta_{23} =$ 0.41(0.59) for LO(HO) and true $\delta_{CP}$= $\pi/2$.}
\label{t23dcp=90}
\end{figure}

Another way to understand the degeneracies among the oscillation parameters by looking at  $\sin^2\theta_{23}$ - $\delta_{CP}$ plane. 
In this section, we  show the $1\sigma$, $2\sigma$, and 90\% C.L. regions for $\sin^2\theta_{23}$ vs. $\delta_{CP}$ for both NO$\nu$A(2+2) and 
NO$\nu$A(3+1). Fig. \ref{t23dcp=0} shows the C.L. regions for NO$\nu$A with true $\sin^2\theta_{23}$ = 0.41 (0.59) for LO (HO) and true $\delta_{CP}$= 0, 
whereas Fig. \ref{t23dcp=90} corresponds to true $\delta_{CP}$= $\pi/2$. Further, the true hierarchy is assumed to be Normal Hierarchy and the C.L. 
regions are obtained both for correct hierarchy (NH-LO and NH-HO) and wrong hierarchy (IH-LO and IH-HO) combinations. 
The black dots in these figures correspond to the assumed true values. From these figures, we can see that NO$\nu$A(2+2) has better degeneracy 
discrimination capability than that of NO$\nu$A(3+1). 

\section{Summary and Conclusions}

At this point of time, where NO$\nu$A experiment already started taking data, it is crucial to analyze how to extract the  best results 
from this experiment with shortest time span for a complete understanding of oscillation parameters. In this paper, we discussed the physics 
potential as well as the role of parameter degeneracies in the determination of oscillation parameters of NO$\nu$A experiment with a total of 
four years of runs with (2$\nu$+2$\bar\nu$) mode.  We find that the  parameter degeneracy discrimination capability  of NO$\nu$A (2+2)  is quite good 
when compared with NO$\nu$A (3+1). Looking all these results from our analysis, it is strongly urged that after two years of neutrino running,
NO$\nu$A should run for two years in antineutrino mode to provide better information about the determination of neutrino  mass ordering and the
octant of atmospheric mixing angle.   

{\bf Acknowledgments}
We would like to thank Science and Engineering Research Board (SERB),
Government of India for financial support through grant No. SB/S2/HEP-017/2013.

\end{document}